\documentclass[fleqn,usenatbib]{mnras}

\usepackage[T1]{fontenc}

\DeclareRobustCommand{\VAN}[3]{#2}
\let\VANthebibliography\thebibliography
\def\thebibliography{\DeclareRobustCommand{\VAN}[3]{##3}\VANthebibliography}

\usepackage{graphicx}	
\usepackage{amsmath}	
\usepackage{amssymb}	
\usepackage{pdflscape}
\usepackage{mdframed}  
\usepackage{float}      
\usepackage{mathtools}
\usepackage{hyperref}
\usepackage{multirow}
\usepackage{multicol}
\usepackage{tcolorbox}
\usepackage{adjustbox}
\usepackage{booktabs}
\usepackage[para, flushleft]{threeparttable}
\hypersetup{colorlinks,linkcolor={blue},citecolor={blue},urlcolor={red}}  
\usepackage{newtxtext,newtxmath}
\usepackage{orcidlink}

\DeclareMathOperator{\Ha}{H\alpha}
\DeclareMathOperator{\QHa}{\textit{Q}(H\alpha)}
\DeclareMathOperator{\LHa}{\textit{L}_{H\alpha}}
\DeclareMathOperator{\HII}{H~\textsc{ii}}
\DeclareMathOperator{\f}{\textit{f}_{\textrm{esc}}}
\DeclareMathOperator{\fDIG}{\textit{f}_{\textrm{DIG}}}
\DeclareMathOperator{\fphot}{\textit{F}_{phot}}
\DeclareMathOperator{\QH}{\textit{Q}(H^0)}
\DeclareMathOperator{\Mo}{M_\odot}
\DeclarePairedDelimiter\autobracket{(}{)}
\newcommand{\br}[1]{\autobracket*{#1}}

\usepackage{newtxtext,newtxmath}

\newcommand{\aref}[1]{\hyperref[#1]{Appendix~\ref{#1}}}

\title[Escape fraction of NGC 628 with LEGUS-SIGNALS]{Constraining the LyC escape fraction from LEGUS star clusters with SIGNALS $\HII$ region observations: A pilot study of NGC 628}

\author[Teh et al.]{Jia Wei Teh\orcidlink{0000-0001-7863-5047}$^{1,2,3}$\thanks{\href{mailto:jiawei.teh@stud.uni-heidelberg.de}{jiawei.teh@stud.uni-heidelberg.de}}, 
Kathryn Grasha\orcidlink{0000-0002-3247-5321}$^{2,3,27}$\thanks{ARC DECRA Fellow}, 
Mark R.\ Krumholz\orcidlink{0000-0003-3893-854X}$^{2,3}$, 
Andrew J.\ Battisti\orcidlink{0000-0003-4569-2285}$^{2,3}$, 
Daniela Calzetti\orcidlink{0000-0002-5189-8004}$^{4}$,
\newauthor 
Laurie Rousseau-Nepton\orcidlink{0000-0002-5136-6673}$^{5,6}$,
Carter Rhea\orcidlink{0000-0003-2001-1076}$^{7,8}$,
Angela Adamo\orcidlink{0000-0002-8192-8091}$^{9}$,
Robert C.\ Kennicutt$^{10,11}$,
Eva K.\ Grebel\orcidlink{0000-0002-1891-3794}$^{12}$,
\newauthor 
David O.\ Cook\orcidlink{0000-0002-6877-7655}$^{13}$,
Francoise Combes\orcidlink{0000-0003-2658-7893}$^{14}$,
Matteo Messa\orcidlink{0000-0003-1427-2456}$^{9,15}$,
Sean T.\ Linden\orcidlink{0000-0002-1000-6081}$^{4}$,
Ralf S.\ Klessen\orcidlink{0000-0002-0560-3172}$^{1,16}$,
\newauthor 
Jos\'{e} M.\ Vilchez\orcidlink{0000-0001-7299-8373}$^{17}$,
Michele Fumagalli\orcidlink{0000-0001-6676-3842}$^{18,19}$,
Anna McLeod\orcidlink{0000-0002-5456-523X}$^{20,21}$,
Linda J.\ Smith\orcidlink{0000-0002-0806-168X}$^{22}$,
Laurent Chemin\orcidlink{0000-0002-3834-7937}$^{23}$,
\newauthor 
Junfeng Wang\orcidlink{0000-0003-4874-0369}$^{24}$,
Elena Sabbi\orcidlink{0000-0003-2954-7643}$^{22}$,
Elena Sacchi\orcidlink{0000-0001-5618-0109}$^{25}$,
Andreea Petric\orcidlink{0000-0003-4030-3455}$^{22}$,
Lorenza Della Bruna$^{9}$,
\newauthor 
Alessandro Boselli\orcidlink{0000-0002-9795-6433}$^{26}$
\\\\
\textit{Affiliations are listed after the references.}
}

\pubyear{2022}

\begin{document}
\label{firstpage}
\pagerange{\pageref{firstpage}--\pageref{lastpage}}
\maketitle

\begin{abstract}
The ionising radiation of young and massive stars is a crucial form of stellar feedback. Most ionising (Lyman-continuum; LyC, $\lambda < 912$\AA) photons are absorbed close to the stars that produce them, forming compact $\HII$ regions, but some escape into the wider galaxy. Quantifying the fraction of LyC photons that escape is an open problem. In this work, we present a semi-novel method to estimate the escape fraction by combining broadband photometry of star clusters from the Legacy ExtraGalactic UV Survey (LEGUS) with $\HII$ regions observed by the Star formation, Ionized gas, and Nebular Abundances Legacy Survey (SIGNALS) in the nearby spiral galaxy NGC 628.  We first assess the completeness of the combined catalogue, and find that 49\% of $\HII$ regions lack corresponding star clusters as a result of a difference in the sensitivities of the LEGUS and SIGNALS surveys. For $\HII$ regions that do have matching clusters, we infer the escape fraction from the difference between the ionising power required to produce the observed $\HII$ luminosity and the predicted ionising photon output of their host star clusters; the latter is computed using a combination of LEGUS photometric observations and a stochastic stellar population synthesis code \textsc{slug} (Stochastically Lighting Up Galaxies). Overall, we find an escape fraction of $\f = 0.09^{+0.06}_{-0.06}$ across our sample of 42 $\HII$ regions; in particular, we find $\HII$ regions with high $\f$ are predominantly regions with low $\Ha$-luminosity. We also report possible correlation between $\f$ and the emission lines $\rm[O\,\textsc{ii}]/[N\,\textsc{ii}]$ and $\rm[O\,\textsc{ii}]/H\beta$.
\end{abstract}




\section{Introduction}
\label{sec:introduction}

Stars are formed primarily in groupings called clusters \citep[e.g., see][]{portegies2010young, krumholz2019star}. Observations of star clusters provide information not just on the properties of clusters themselves, such as the cluster mass and age function \citep{larsen2009mass, anders2021star}, but also allow us to study the hierarchical structure of star-forming galaxies \citep[e.g.,][]{2001ApJ...561..727Z,2005A&A...443...79B, 2017ApJ...840..113G} and to understand the history of star formation \citep[e.g.,][]{2011ApJ...741..108P, 2013MNRAS.430..676B}. In particular, young star clusters \citep[YSCs; $t \lesssim$ 10 Myr; see][]{2008A&A...489.1091M,portegies2010young, 2016ApJ...817....4F} are very luminous, and can easily be detected and used to trace active star-forming regions in galaxies that are too distant for individual stars to be resolved \citep[e.g.,][]{1999AJ....118.1551W, 2005A&A...443...79B, 2009ApJ...701.1015K, 2010ApJ...725.1620A, 2011AJ....142...42F, 2017ApJ...842...25G, 2022A&A...662L...6B}. One particularly important application of observations of YSCs is to constrain how massive ($\gtrsim$ 8 $\Mo$) stars drive feedback into the surrounding interstellar medium (ISM; for a review see \citealt{krumholz2019star}), a process that remains only partially understood at best. In this paper, we focus on one aspect of that feedback: ionising radiation. Lyman-continuum (LyC; $\lambda < 912$\AA) emission from O- and B-type stars ionises the ISM around them, forming $\HII$ regions. However, some fraction of the LyC flux may escape the $\HII$ region and contribute to ionisation elsewhere in the diffuse medium of the galaxy \citep{2016A&A...592A..47N}. Quantifying the LyC escape fraction, $\f$, thus allows us to probe the origin of ionisation of the diffuse ionised gas (DIG) in galaxies \citep[e.g., see][]{2000ApJ...541..597H, 2002A&A...386..801Z, 2007ApJ...661..801O, 2009ApJ...703.1159S, 2018A&A...611A..95W, 2021A&A...650A.103D, 2022A&A...659A..26B}. Some fraction of the LyC flux that escapes $\HII$ regions may also escape the galaxy entirely and propagate into the intergalactic medium; constraining the fraction that does so is important to the study of cosmological reionisation \citep[e.g., see][]{2011A&A...530A..87P, 2013MNRAS.428L...1M, 2017MNRAS.468..389J, 2020A&A...644A..21R, 2022MNRAS.517.5104C, 2022A&A...663A..59S, 2022ApJS..260....1F, 2022ApJ...930..126F}. Thus, the measurements of local $\f$ from star-forming regions impose an upper limit on the latter process.

Astronomers have used a variety of techniques to compute the LyC escape fraction, though there has not been a strong constraint on the range of $\f$. One common method is to use surveys of individual massive stars to evaluate the LyC flux, and compare directly to the observed $\Ha$ luminosity of their $\HII$ regions \citep[e.g.,][]{1997MNRAS.291..827O,2013A&A...558A.134D, 2019MNRAS.486.5263M, 2020ApJ...891...25M, 2022PASP..134f4301G}. Recently, \citet{2019MNRAS.486.5263M} combined MUSE (Multi-Unit Spectroscopic Explorer) observations of 11 $\HII$ regions in the LMC with spectroscopy of massive O stars to directly link the massive stellar population and feedback-related quantities of the regions, yielding a mean escape fraction of $\f \sim 0.45$. In subsequent work, \citet{2020ApJ...891...25M} combined observations from MUSE with spectra of massive stars to study two star-forming complexes of the nearby dwarf spiral galaxy NGC 300 and found $\f \sim 0.28$ and $\f \sim 0.51$. More recently, \citet{2020ApJ...902...54C} modelled the spectral energy distribution of resolved stars in NGC~4214 and find substantial variation in the LyC escape fraction ($0\%-40\%$). While these surveys represent a significant advance, further progress can be made in two directions: first, with the exception of nearby galaxies, traditional methods of determining the LyC escape fraction are ineffective in extragalactic surveys where spectra for individual massive stars are difficult to obtain. Second, the observations are limited by the small field-of-view (FoV) of these instruments; a larger and more statistically significant sample is necessary to ensure that these results are robust.

The difficulty of obtaining spectra in extragalactic surveys has led to methods to estimate stellar LyC fluxes, and subsequently escape fractions, using broadband photometry rather than spectroscopy. \citet{2016A&A...592A..47N} evaluated the spectral types of the most massive stars in NGC 300 using broadband data from the \textit{HST} and stellar atmosphere models, then compared with surrounding $\Ha$ observations. However, they found that the resulting escape fraction is completely dominated by uncertainties in determining stellar parameters from photometric data. \citet{2018A&A...611A..95W} computed the LyC escape fraction of $\HII$ regions in the Antennae galaxy by comparing $\HII$ region luminosities from MUSE observations to LyC emission from star clusters derived from broadband photometry from \textit{HST}. They estimated the LyC flux by comparing the measured photometry to the GALEV evolutionary synthesis models \citep{2009MNRAS.396..462K} and \textsc{starburst99} models \citep{1999ApJS..123....3L}. However, they lacked photometric data in the UV band, potentially causing small uncertainty in the age estimates of the youngest stellar populations, which often have the strongest LyC flux. Furthermore, the stochasticity of the stellar initial mass function (IMF) sampling was not taken into account when deriving the LyC flux from star clusters, which may affect the derived ionising luminosity of clusters at the low-mass end.

In this paper, we revisit the question of the ionising escape fraction from star clusters using an improved method. First, to ensure that our analysis includes a large enough sample, we use $\HII$ region data in NGC~628 drawn from the survey SIGNALS \citep[The Star formation, Ionized gas, and Nebular Abundances Legacy Survey;][]{2019MNRAS.489.5530R}. NGC~628 is the first galaxy observed by SIGNALS, which uses the SITELLE Imaging Fourier Transform Spectrograph (IFTS) to achieve high spectral resolution over a FoV large enough (11' $\times$ 11') to enable mapping of full galactic discs. The SIGNALS catalogue for NGC~628 includes 4285 $\HII$ regions imaged at a spatial resolution of 35~pc \citep{2018MNRAS.477.4152R}. We combine this catalogue with the LEGUS \citep[Legacy ExtraGalactic UV Survey;][]{2015AJ....149...51C} survey, which provides a total of 1648 candidate star clusters in NGC~628 \citep[see][]{2015ApJ...815...93G, 2017ApJ...841..131A}. Combining these catalogues allows us to study the physical connections between $\HII$ regions and star clusters for an unprecedented sample size and spatial resolution in a spiral galaxy at a distance of $\sim$ 10~Mpc. Second, we adopt a fully stochastic treatment of star cluster LyC fluxes, following the approach of \citet{2020A&A...635A.134D, 2021A&A...650A.103D, 2022A&A...660A..77D, 2022A&A...666A..29D}. We use a combination of five-colour photometric data from \textit{HST} (including the crucial UV band) and a stochastic stellar population code \citep[\textsc{slug};][]{2012ApJ...745..145D, 2014MNRAS.444.3275D, 2015MNRAS.452.1447K} to infer the ionising luminosities $\QH$ from star clusters in NGC~628, and compare this to the observed $\Ha$ luminosity $\LHa$ of their $\HII$ regions. 

Compared to previous works, our study has the advantage of both providing a much larger sample size, and including a treatment of the effects of stochastic IMF sampling. Our approach is similar to that of \citet{2022A&A...666A..29D}: they combined MUSE observations of ionised gas with YSC observations from \textit{HST}, identifying $\sim 4700~\HII$ region samples in the nearby spiral galaxy M83 to study the link between the feedback of young clusters and their surrounding gas. The consideration of the stochastic sampling of IMF is essential for NGC 628, which has low mass ionising star clusters ($\gtrsim 10^{2.5}\Mo$; \citealp[see][]{2015ApJ...815...93G}). If this method is shown to be feasible, then it can be applied to other galaxies in future LEGUS--SIGNALS observations, allowing us to study star formation across a wide range of galactic environments. It can also be implemented in other surveys, such as the PHANGS MUSE nebular \citep{2022A&A...659A.191E} and \textit{HST} star cluster catalogues \citep{2022MNRAS.509.4094T, 2022ApJS..258...10L}.

In addition, the study of correlations between emission line ratios and the escape fraction of LyC photons has been widely explored at \textit{galactic} scales \citep[e.g., see][]{2013ApJ...766...91J, 2014MNRAS.442..900N,2018MNRAS.478.4851I,2019ApJ...885...96J,2020ApJ...889..161N, 2020A&A...644A..21R, 2022ApJS..260....1F, 2022ApJ...930..126F}; at sub-kpc scales, such relations remain partially understood at best. The SIGNALS catalogue, which includes a large number of spatially-resolved optical line luminosity measurements, offers a golden opportunity to investigate this question for all rest-frame optical strong line-ratios: $\rm[N\,\textsc{ii}]/\Ha,
[S\,\textsc{ii}]/\Ha,
[S\,\textsc{ii}]/[N\,\textsc{ii}],
[O\,\textsc{iii}]/H\beta, 
[O\,\textsc{ii}]/H\beta,
[O\,\textsc{iii}]/\allowbreak[O\,\textsc{ii}],
[O\,\textsc{iii}]/[N\,\textsc{ii}],
[S\,\textsc{ii}]\lambda6716/[S\,\textsc{ii}]\lambda6731,$ and
$\rm[O\,\textsc{ii}]/[N\,\textsc{ii}]$.

The remainder of this paper is as follows. We first describe the observational catalogues we use, and our method for combining them, in \autoref{sec:data_selection}. We assess the completeness of our catalogue and evaluate potential biases in \autoref{sec:completeness}. We analyse the data in \autoref{sec:escape_fraction}, discuss the implications in \autoref{sec:discussion} and we summarise our conclusions in \autoref{sec: conclusion}.

\section{Data Selection}
\label{sec:data_selection}

\subsection{NGC 628 overview}
\label{sec: NGC 628}
NGC~628 is a star-forming, grand-design SA(s)c spiral galaxy located at a distance of 9.9 Mpc \citep{2010ApJ...715..833O, 2021MNRAS.501.3621A}.  We select this galaxy for two main reasons. First, NGC~628 is face-on and has a large apparent radius (5.23 $\pm$ 0.24 arcmin; \citealt{2014MNRAS.437.1337G}). These features allow detailed study of star-forming regions (e.g., \citealp{2006ApJ...644..879E,2015ApJ...806...16B, 2018ApJ...863L..21K,2021MNRAS.506.5294W, 2022MNRAS.509.4094T, 2022ApJS..258...10L}; Scheuermann et al. in prep.), dust and metal content \citep{2019MNRAS.483.4968V}, and the temporal evolution of cluster sizes \citep{2015ApJ...815...93G, 2017ApJ...841...92R}. Second, NGC~628 is the first galaxy to be observed by both the SIGNALS and LEGUS surveys, providing a large sample of star clusters and $\HII$ regions for analysis.

\subsection{Star cluster data}
\label{sec: starclusters}

The catalogue of star clusters we use in this work comes from the LEGUS survey \citep[see][]{2015ApJ...815...93G, 2017ApJ...841..131A}. LEGUS  \citep{2015AJ....149...51C} is a Cycle 21 Treasury program on the \textit{HST} targeting 50 local ($\lesssim$ 12 Mpc) galaxies. The observations provide broad-band imaging in the NUV (F275W), U (F336W), B (F438W), V (F555W), and I (F814W) filters, covered either by the Wide Field Camera 3 (WFC3) or the Advanced Camera for Surveys (ACS). The star cluster catalogues of LEGUS have enabled a broad range of star formation studies, including testing bar-driven spiral density wave theory \citep{2018MNRAS.478.3590S}, the nature and the spatial and temporal evolution of the hierarchical structures of star clusters \citep{2015ApJ...815...93G, 2017ApJ...840..113G, 2017ApJ...842...25G, 2021MNRAS.507.5542M, 2021MNRAS.506.5294W, 2022ApJS..258...10L}, IMF studies \citep{2015ApJ...812..147K, 2017MNRAS.469.2464A}, timescales for star clusters to disassociate with their parent GMCs \citep{2018MNRAS.481.1016G, 2019MNRAS.483.4707G}, and $\HII$ region evolution timescales \citep{2019MNRAS.490.4648H, 2020ApJ...889..154W}.

LEGUS observations of NGC~628 consist of an east pointing (NGC628e) and a centre pointing (NGC~628c). At the distance of NGC~628, 9.9~Mpc, the pixel size of \textit{HST}, $0\farcs 04$ corresponds to a spatial resolution of 1.9~pc. Detailed descriptions of the survey and the standard data reduction of LEGUS imaging datasets are available in \citet{2015AJ....149...51C}, whereas the custom pipelines developed to create the initial cluster candidate catalogues, and the analysis of cluster population (i.e., cluster mass function, disruption timescales) are presented in \citet{2017ApJ...841..131A}. Here, we provide a brief description of steps taken in the automatic pipelines to produce the catalogue of star clusters used in this work. First, the SExtractor \citep{1996A&AS..117..393B} algorithm is used on white-light (combination of all five filters) images to extract objects with at least a 3$\sigma$ detection in a minimum of 5 contiguous pixels. Next, candidates must have a $V$ band concentration index (CI; as the difference in magnitudes between radii of 1 pixel and 3 pixels) $>1.4$ mag for the centre pointing, and $>1.3$ mag for the east pointing, and must be detected in the V band and either B or I band, and a total of at least four bands, with photometric error $\sigma_\lambda\leq 0.3$ mag \citep[see][for justifications of cluster selection criteria]{2017ApJ...841..131A}. Second, each cluster candidate with an absolute V band magnitude $<-6$ mag in the automatic catalogue is then visually inspected and assigned one of four classes: (1) centrally concentrated clusters with spherically symmetric radial profiles; (2) clusters with asymmetric radial profiles; (3) clusters with multiple peaks; and (4) non-clusters such as background galaxies, stars, bad pixels, or chip edge artefacts.

\begin{figure*}
    \centering
    \includegraphics[width = .8\textwidth]{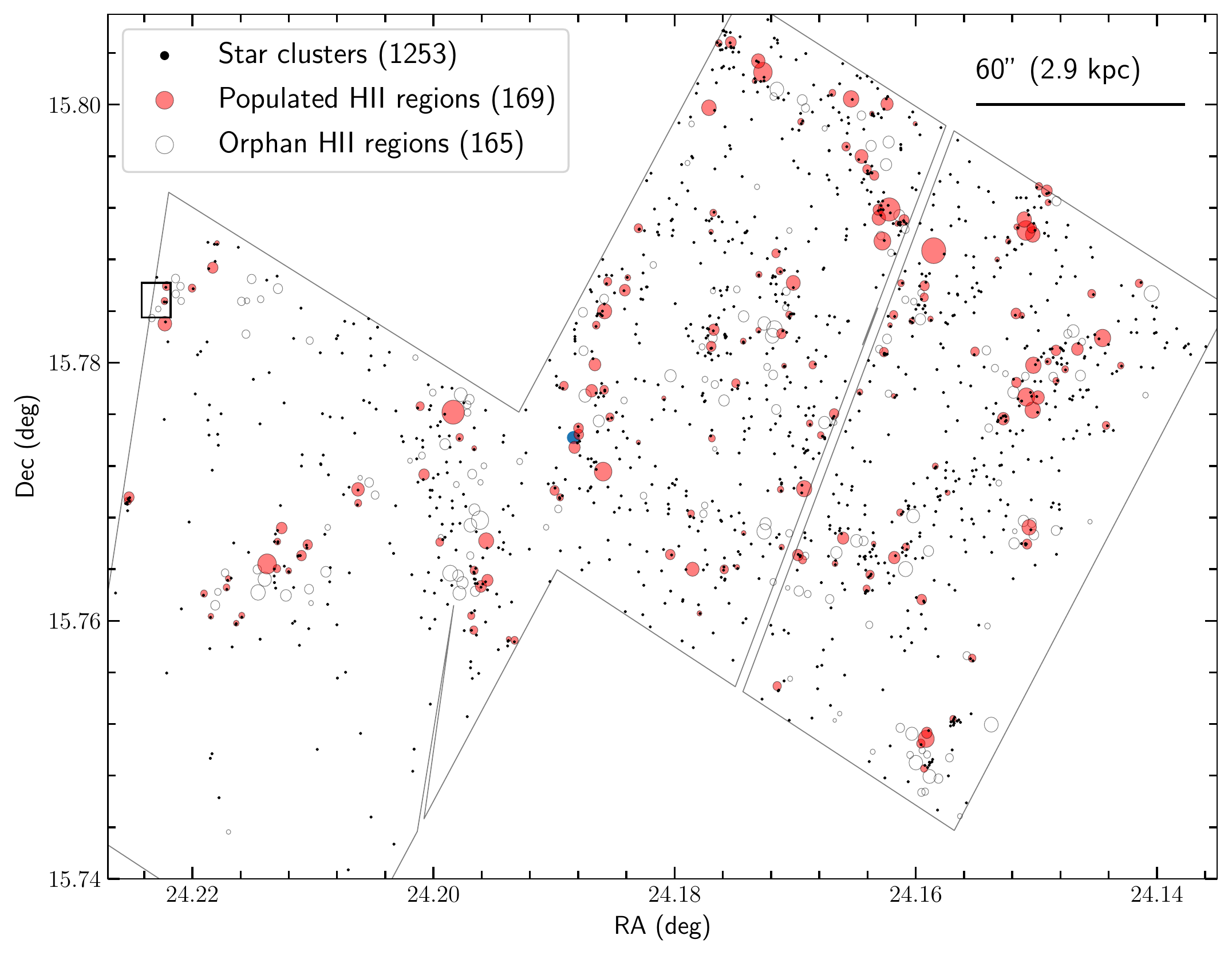}
    \caption{Spatial distribution of LEGUS classified star clusters (Class~1, 2, and 3; in black dots) and symmetrical + asymmetrical $\HII$ regions (in circles, showing true physical sizes with respect to the axes ticks). The filled (red) and empty circles show associated and orphan $\HII$ regions respectively. The number of samples in each category is given in the parenthesis. The black outline represents the \textit{HST} FoV defined in \autoref{sec: starclusters}. The black horizontal bar indicates a $60''$ length, corresponding to 2.9~kpc at distance of NGC~628. A cutout box is randomly selected (black box) for visual inspection (see \autoref{fig: resolution_check}).}
    \label{fig: Catalogue}
\end{figure*}

In most of this work, we limit our analysis to Class~1, 2, and 3 clusters, since these are verified to be genuine clusters (or compact associations) by the LEGUS team, using a combination of visual inspection and machine learning techniques \citep{2019MNRAS.483.4707G}. We make an exception to this in \autoref{sec: class4} and \autoref{sec: individual} for reasons we discuss in \autoref{sec:completeness}, but we set aside the Class 4 sources for the moment. Overall, the LEGUS catalogue contains 1253 Class~1, 2 and 3 clusters. The effective FoV of the survey consists of the regions where there is enough overlapping coverage of the various LEGUS filters such that it is possible to categorise sources into Class~1 -- 4, which means that there must be overlapping coverage of: (1) V; (2) B or I; and (3) a total at least 4 bands. We show the footprint of the region satisfying this constraint in \autoref{fig: Catalogue}.

\begin{figure*}
    \centering
    \includegraphics[width = .8\textwidth]{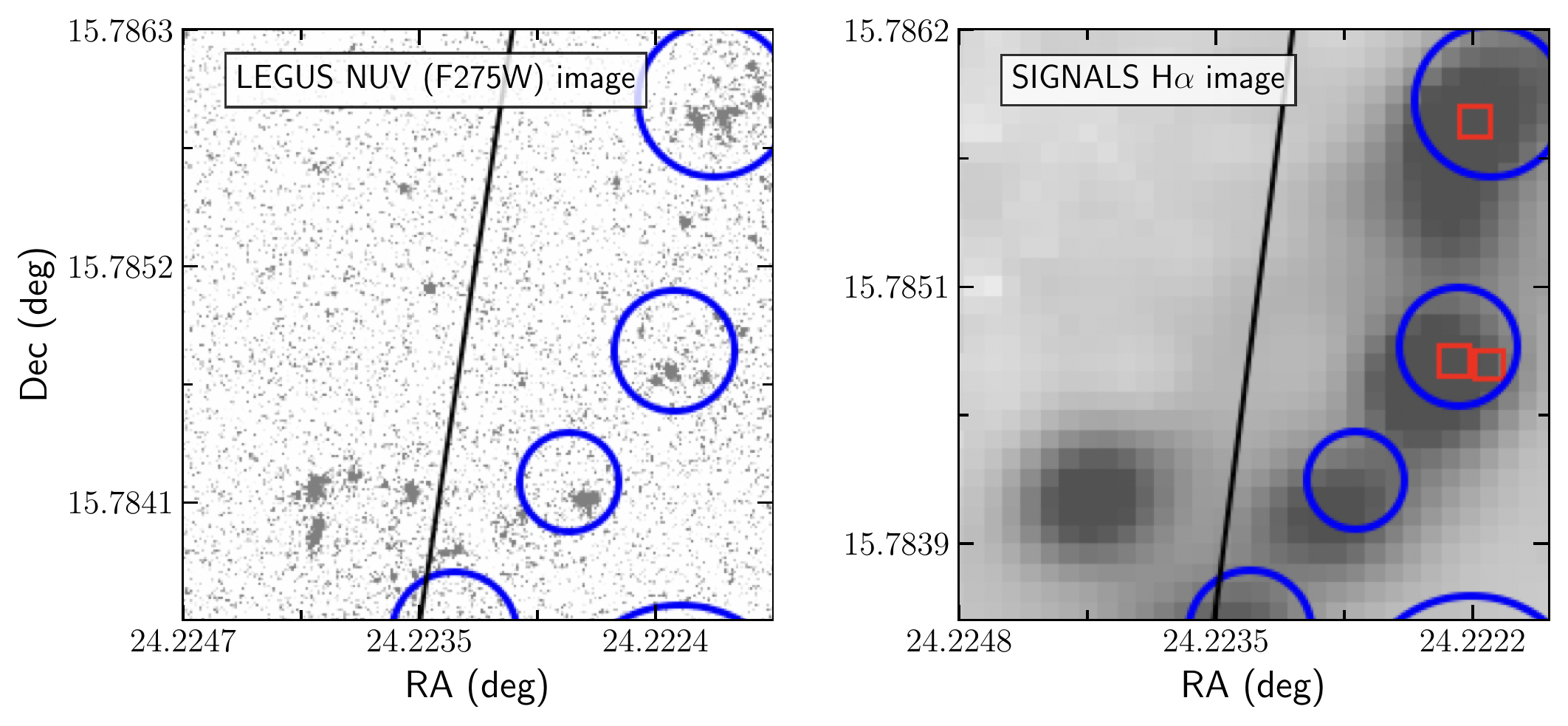}
    \caption{Spatial distribution of $\HII$ regions (blue circles) and Class~1, 2, and 3 sources (red boxes) in the cutout defined in \autoref{fig: Catalogue}, overlaid on the NUV (F275W) image from the LEGUS survey (left panel) and the $\Ha$ image from the SIGNALS survey (right panel). We show the red cluster markers only in the right panel to avoid covering up the bright points to which they correspond in the NUV image. Black lines indicate the edge of the LEGUS FoV shown in \autoref{sec: starclusters}.}
    \label{fig: resolution_check}
\end{figure*}

The LEGUS catalogue also assigns an age, mass, and extinction to every cluster using \texttt{Yggdrasil} \citep{2011ApJ...740...13Z} deterministic stellar population models, using a $\chi^2$-fitting procedure described in \citet{2017ApJ...841..131A}, including propagation of photometric uncertainties through the fit \citep[see][]{2010ApJ...725.1620A}. For NGC~628, the fits assume Solar metallicity in both gas and stars, and include nebular emission computed assuming a 50\% covering fraction, a Milky Way extinction curve \citep{1989ApJ...345..245C}, and a foreground extinction of $E(B-V) = 0.06$.

\subsection{H~\textsc{ii} region data}
\label{sec: hii}

The $\HII$ region catalogue comes from the SIGNALS survey \citep{2019MNRAS.489.5530R}. SIGNALS is ongoing, and will eventually observe more than 50\,000 $\HII$ regions in $\sim$ 40 nearby galaxies, using three spectral filters, SN1, SN2, and SN3, covering important optical emission lines for detailed chemical studies of the ISM \citep[see][]{2019MNRAS.489.5530R}. SIGNALS has a spatial resolution of $\approx 0\farcs 8$, corresponding to 35~pc at the distance of NGC~628 \citep[e.g., see][]{2020ApJ...901..152R, 2021ApJ...910..129R, 2021ApJ...923..169R}. Detailed descriptions of data reduction are available in \citet{2018MNRAS.477.4152R}. The development of artificial neural network techniques will allow for robust and efficient estimates of the kinematic parameters for the optical emission-line spectra of all the $\HII$ regions within SIGNALS.

In SIGNALS, an $\HII$ region is defined via a watershed algorithm: one first identifies an emission peak, then determines a zone of influence around it, and lastly outlines an outer limit where the region merges into the DIG background. Once defined, the intensity profile of each $\HII$ region is fit to a pseudo-Voigt function, and based on this fit it is assigned one of four categories: (1) symmetric profile, with central concentrated ionising sources; (2) asymmetric profile, with dispersed ionising sources; (3) transient region; and (4) diffuse region \citep[see][]{2019MNRAS.489.5530R}. We note that the $\HII$ regions in the SIGNALS catalogue are reported as circular regions, by transposing the pseudo-Voigt fitted profile half-width $\sigma$ into radius \citep[see][for more details]{2018MNRAS.477.4152R}; we discuss the impact of this approximation in \autoref{sec: astrometric}. 

Additional measures were taken to minimise the inclusion of non-$\HII$ regions (e.g., DIG regions, supernova remnants, and planetary nebulae) in the SIGNALS catalogue \citep{2018MNRAS.477.4152R}. Nevertheless, we choose a subset of $\HII$ regions by applying two additional constraints to further strengthen the fidelity of our sample regions. First, we only select concentrated symmetrical and asymmetrical $\HII$ regions in the FoV for the analysis, as transient and diffuse $\HII$ regions have intensity profiles that are highly dispersed and are likely dominated by DIG. Second, we check the location of our selected regions within a BPT diagram \citep{1981PASP...93....5B, 2006MNRAS.372..961K} of $\rm[O\,\textsc{iii}]/H\beta$ vs $\rm[N\,\textsc{ii}]/\Ha$ and $\rm[O\,\textsc{iii}]/H\beta$ vs $\rm[S\,\textsc{ii}]/\Ha$, and retain only regions that lie within the star-forming regime. These constraints reduce the number of $\HII$ region candidates from 4067 to 334; we further discuss the possible effect of this selection criteria on our analysis in \autoref{sec: fesc_implication}.

\subsection{Constructing the combined catalogue}
\label{sec: combined}

Given our set of 334 $\HII$ regions with a specified radius $R$, and 1253 star clusters, each with a specified central position, we refer to a cluster and $\HII$ region as associated if the projected distance $D$ between the central positions is $\leq R$. While this geometrical approach provides a reasonable definition of overlapping regions, we pause to point out three caveats. First, we are treating star clusters as point sources. The approximation here is small, since most clusters have radii $\sim$ 3~pc \citep{2017ApJ...841...92R, 2021MNRAS.508.5935B}, $\sim$ 1 order of magnitude smaller than the SIGNALS resolution of 35~pc. Second, we can only assess overlap in projection, as we do not have information for the depth dimension. This means that false associations due to chance alignments in the line-of-sight are inevitable; we discuss the impact of these in \autoref{sec: individual}. Third, we do not impose an age cutoff on star clusters in the process of association, for reasons we will also discuss in \autoref{sec: individual}; this too almost certainly contributes to false associations due to chance overlaps

We present the combined census of 1253 star clusters, and 334 $\HII$ regions in \autoref{fig: Catalogue}. In \autoref{fig: resolution_check} we zoom in on a small randomly-selected portion of the field, showing the LEGUS NUV and SIGNALS H$\alpha$ maps directly. It is clear that in cases where there are cluster-H~\textsc{ii} region overlaps, the association is relatively unambiguous: many of the blue circles marking SIGNALS-catalogued H~\textsc{ii} regions have one or two obvious slightly-extended NUV sources near their centres. Conversely, in the H$\alpha$ image we clearly see that many of the H$\alpha$ maxima have clusters sitting on top of them. However, we also observe that many of the $\HII$ regions do not seem to host star clusters -- 165 out of 334 (49\%) $\HII$ regions are ``orphans'' (refer to \autoref{tab: terminology} for terminology). We investigate this topic next.

\begin{table*}
	\centering
	\caption{A list of the terminology and symbols used throughout this work.}
	\label{tab: terminology}
	\begin{tabular}{ll} 
		\hline
		\hline
		Terminology/Symbols & Description   \\
		\hline
		Orphan $\HII$ regions & $\HII$ regions that lack ionising star clusters. \\
		Populated $\HII$ regions & $\HII$ regions that contain ionising star clusters. \\
		$\QH \text{ (photons s}^{-1})$ & The computed LyC flux from star clusters using LEGUS photometric observation.\\
		$\LHa$ (erg s$^{-1}$)& The $\Ha$ luminosity of $\HII$ regions from SIGNALS observation.\\
		$\QHa \text{ (photons s}^{-1})$ & The LyC flux corresponding to $\LHa$, 
		 obtained with $\QHa = 7.31\times10^{11} \LHa$ \citep{1998ApJ...498..541K}. \\
        \hline
	\end{tabular}
\end{table*}

\section{Catalogue Completeness}
\label{sec:completeness}
The phenomenon of $\HII$ regions without detected star clusters is particularly intriguing, as there must be an ionising source that is injecting photons into these regions. How could so many ionising sources of these $\HII$ regions have been missed? In this section, we present six possible explanations for the high fraction of orphan $\HII$ regions. Such an analysis is necessary because, unlike in prior work using targeted observations of $\HII$ regions around particular, known star clusters, we are working with catalogues of clusters and $\HII$ regions that were made independently of one another. If we are to use the combined catalogue to assess escape fractions, we must understand if there are systematic errors or biases in the two source catalogues we are combining.

\subsection{Astrometric uncertainties}
\label{sec: astrometric}

We begin by considering the effect of astrometric uncertainties. If the astrometric registration between the LEGUS and SIGNALS images is different, causing an offset in the images, then some $\HII$ regions that actually contain star clusters might be misclassified as orphans.

For the SIGNALS survey, a detailed description of the astrometric calibration for all filters in SITELLE is presented in \citet{2019MNRAS.489.5530R}. They used bright stars in the FoV to calculate the astrometric solution and showed that the resulting calibration has less than $0\farcs 324$ uncertainty ($\sim 1$ SITELLE pixel) almost everywhere throughout the FoV. An additional calibration for the SN3 filter in SITELLE is presented in \cite{2018MNRAS.473.4130M}, showing that the results are accurate to $< 1$ pixel. \cite{2018MNRAS.473.4130M} also compared SITELLE's catalogue of emission-line point-like sources in M31 to other catalogues \citep[][]{2006MNRAS.369...97H,2006MNRAS.369..120M} and found an upper-limit uncertainty of $0\farcs 21$. 

In the LEGUS survey, the WFC3/UVIS data were first aligned using \texttt{TWEAKREG} routine in \texttt{drizzlepac} \citep{2012drzp.book.....G}. The shifts, scale and rotation of individual exposures were then solved and used for matching with catalogues typically containing few hundred bright sources for alignment purposes. The accuracy is better than $0\farcs 004$\footnote{We note that the quoted accuracy is limited by the absolute astrometric accuracy of Guide Star II catalogue \citep[$< 0\farcs 4$; ][]{2008AJ....136..735L}. Regardless, this is at most $\lesssim 2$ SITELLE pixels, and will not affect our analysis.}.

These upper limits on the astrometric error are sufficient to rule out astrometric offsets as a significant contributor to the orphan population. Of our 165 orphan $\HII$ regions, only 5 have a star cluster within even $0\farcs 1 (\sim 4.8~\rm{pc})$ of their outer edge (as defined by their assigned radius in SIGNALS). Thus even if we were to shift the positions of the images by the maximum amount allowed by the possible registration errors, the number of orphan $\HII$ regions that could be shifted to the associated category thereby is negligible.

\subsection{Single-star H~\textsc{ii} regions}
\label{sec: singleStar}
A second possible explanation for the orphan $\HII$ regions is that they are ionised by single, isolated O- or B-type star. In LEGUS, single stars are separated from clusters based on their CI (see \autoref{sec: starclusters}), so a single isolated massive star would not be classified as a cluster, even if it had ionising luminosity large enough to create detectable $\HII$ region. However, this too seems unlikely to be a significant contributor to orphans. While $10\%-30\%$ of massive stars are found in isolation from star-forming regions \citep{1987ApJS...64..545G, 2004AJ....127.1632O}, most, perhaps all, of these are runaways \citep[e.g.,][]{2005A&A...437..247D,2007MNRAS.380.1271P,2010MNRAS.404.1564P,2012MNRAS.424.3037G}. Such stars will be found far from their natal gas, in regions of low gas density \citep[e.g.,][]{2001A&A...365...49H, 2018MNRAS.480.2109D}, where the $\HII$ regions they create are likely to be undetectable at extragalactic distances due to low H$\alpha$ surface brightness. Even in cases where these massive stars travel with low velocities \cite[walkaways; e.g., see][]{2012ASPC..465...65D, 2019A&A...624A..66R}, or that they travel along the galactic plane and end up in high-density regions, the $\HII$ regions they produce are typically compact and have sizes $\lesssim 10$ pc \citep[e.g., see][]{2011A&A...530A..57S, 2013MNRAS.436..859M}. Given that the orphan regions in SIGNALS have a median radius of $\sim 56$ pc, this suggests that the isolated stars may not be a major contributor to the observed phenomenon.

\subsection{Over-segmentation}
\label{sec: oversegmentation}

Another possible explanation for the orphan $\HII$ regions comes from the possibility of over-segmentation of $\HII$ regions from the SIGNALS survey. The idea is fairly straightforward: an area consisting of multiple bright regions of $\Ha$ emission, which SIGNALS decomposes into several $\HII$ regions, could in reality be a single, giant $\HII$ region that is ionised by a central YSC.

\citet{2018MNRAS.477.4152R} describes the steps taken to minimize the possibility of false $\HII$ region emission peaks. They imposed an intensity threshold on neighbouring pixels of an emission peak, preserving only the brightest peak if two emission peaks are at a distance smaller than the image quality of the observation. They also varied radii of apertures centred on the emission peak to check if the mean flux decreases as the radii increases. However, these checks only protect against over-segmentation due to surface brightness fluctuations; they are of little help if the $\Ha$ surface brightness really is multiply-peaked, but is nonetheless powered by a single source. This could occur in two possible scenarios. First, multiple nearby emission peaks could correspond to spatially-separate regions of high-density gas ionised by the same star cluster. Second, a foreground absorber such as a dust lane across an $\HII$ region \citep[e.g.,][]{1981PhDT.........3B} could artificially create surface brightness variations in the $\HII$ region large enough for it to be decomposed into multiple emission peaks by the algorithm in SIGNALS. Either of these scenarios could cause an over-segmentation of $\HII$ regions, creating the illusion that these regions are ionised by different individual star clusters. If this phenomenon occurs, then we must not simply assume a one-to-one relation for $\HII$ regions and star clusters, but must also consider orphan $\HII$ regions that are adjacent to a populated $\HII$ region that might share the same ionising source. 

\begin{figure}
    \centering
    \includegraphics[width = \columnwidth]{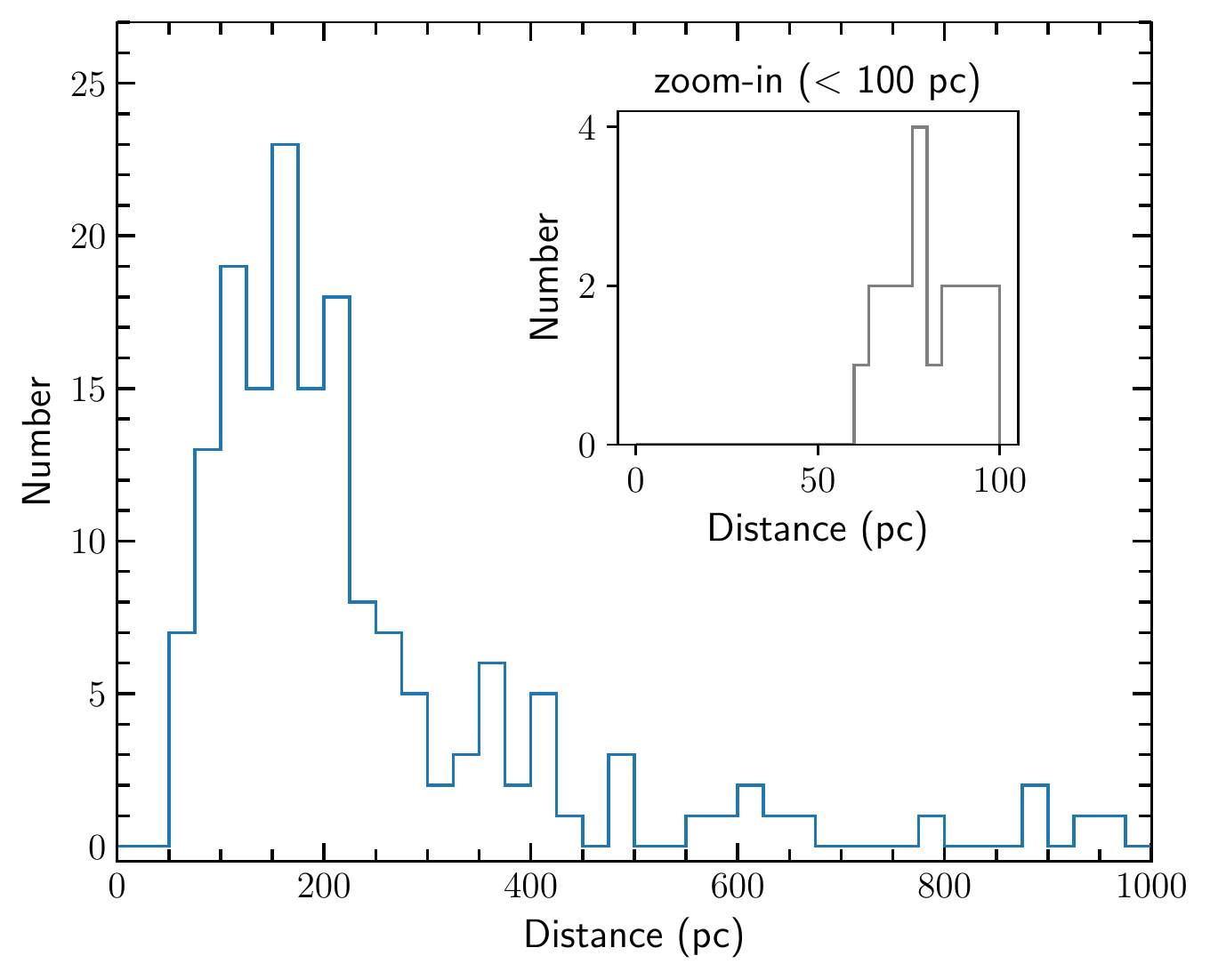}
    \caption{Distance distribution for each orphan $\HII$ region to its nearest populated $\HII$ region. The inset shows a zoom-in on the part of the histogram in smaller bin ($\sim$ 4 pc), showing the number of populated $\HII$ regions within distances of $<100$~pc. We find 20 orphan $\HII$ regions out of 165 have a populated neighbour within 100~pc.}
    \label{fig: Oversegmentation}
\end{figure}

To assess this possibility, in \autoref{fig: Oversegmentation} we show the distribution of distances from each orphan $\HII$ region to its nearest populated $\HII$ region. The plot shows that only 20 orphan $\HII$ regions out of 165 have a populated neighbour within 100~pc. Moreover, 100~pc is a generous estimate for the maximum zone of influence of a star cluster; the median radius of all $\HII$ regions in the SIGNALS catalogue is $\sim$ 74 pc, and for the Str\"omgren radius of an $\HII$ region to reach 100~pc for a typical ionising luminosity of $10^{50}$ erg s$^{-1}$ would require the ionised gas density to be $\lesssim 3\times10^{-3}$ cm$^{-3}$, well below the $\approx 10-100$ cm$^{-3}$ typically inferred from ionised gas density diagnostics in nearby galaxies \citep[e.g.,][]{2019ARA&A..57..511K}. This connotes that over-segmentation of $\HII$ regions, if present, will not be a major cause of orphan $\HII$ regions.

\begin{figure}
    \centering
    \includegraphics[width = \columnwidth]{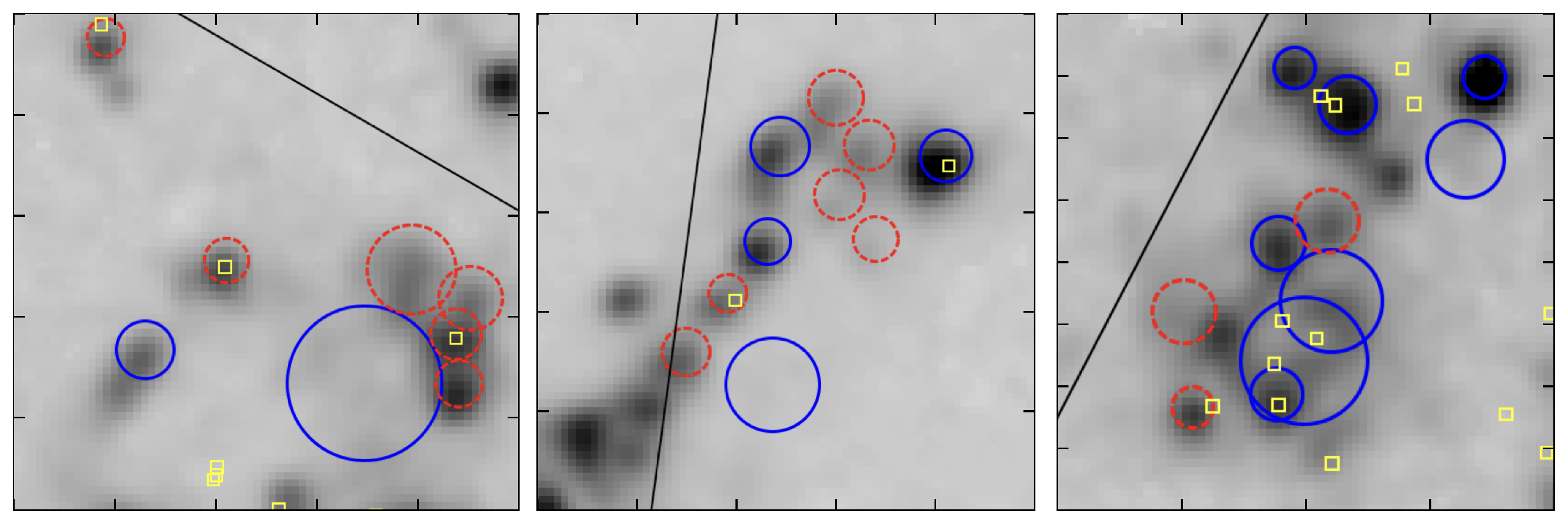}
    \caption{Spatial distribution of Class~4 sources (yellow boxes), orphan $\HII$ regions (red dashed circles), and populated $\HII$ regions (blue circles), overlaid on the $\Ha$ map obtained from the SIGNALS survey. The black outline indicates the FoV defined in \autoref{sec: starclusters}. Class~4 sources are seen associated with both orphan and populated $\HII$ regions. These Class~4 sources and $\HII$ regions are, based on their locations relative to orphan $\HII$ regions, possibly associated with them.}
    \label{fig: class4}
\end{figure}

\subsection{Potential true clusters in Class~4 LEGUS Sources}
\label{sec: class4}

Here, we consider the possibility that the orphan $\HII$ regions are due to star clusters being falsely identified as non-clusters (Class 4). Identification errors of this type are plausible: \citet{2017ApJ...841..131A} performed a comparison between clusters of NGC~628 identified from LEGUS and \citet{2014AJ....147...78W}, and found only $\sim$ 75\% of clusters are common between the studies, as a result of a mix of human classification and automated identification. Additionally, \citet{2019MNRAS.483.4707G} compared classifications in the LEGUS NGC~5194 cluster catalogue and found $70\%-75\%$ agreement between classifiers for the classes. More recently, \citet{2020MNRAS.493.3178W} found that different experts agree among themselves 61\% of the time as to whether a particular candidate in the PHANGS-\textit{HST} NGC~4656 cluster catalogue should be classified as Class 4. 

To assess whether this could contribute to the orphan $\HII$ regions, we overlay Class~4 sources, populated $\HII$ regions and orphan $\HII$ regions on an H$\alpha$ image of NGC~628 from the SIGNALS data. We find that 36 out of 165 orphan $\HII$ regions contain Class~4 sources. Some of these are likely false associations due to chance alignments, but the probability that all 36 associations are chance alignments is very low on both statistical and visual grounds. Statistically, if all Class~4 sources are really background objects or other artefacts, their position should be uniformly distributed across the FoV, and the expected number of chance alignments should then be equal to the fractional area of the FoV covered by orphan $\HII$ regions multiplied by the number of Class~4 sources, which is $\approx 13$; given this expectation value, the probability of finding 36 overlaps is $\sim 10^{-9}$. Visually, at least some of the Class~4 sources are almost perfectly aligned with $\HII$ regions, as we show in \autoref{fig: class4}, which makes chance alignment implausible. Therefore, both visual inspection and statistical analysis suggest at least some Class~4 objects are likely true clusters, and can contribute to the orphan $\HII$ regions; this motivates us to include Class~4 sources in the analysis of $\f$ (\autoref{sec:escape_fraction}). Conversely, however, the fact that we find only 36 Class~4 objects within orphan $\HII$ regions, and that we expect 13 of these to be chance alignments, also suggests that misclassification of true clusters as Class~4 objects in the LEGUS catalogue cannot be a major contributor to the phenomenon of orphan $\HII$ regions.

\subsection{Unclassified star clusters}
\label{sec: unclassified}

Next, we investigate whether the orphan $\HII$ regions could be powered by star clusters too faint for LEGUS to classify. To do so, we create a set of $10^5$ synthetic star clusters using the stochastic stellar population synthesis code \textsc{slug} \citep[Stochastically Lighting Up Galaxies;][]{2012ApJ...745..145D, 2014MNRAS.444.3275D, 2015MNRAS.452.1447K}. We then check what fraction of these star clusters would have ionising luminosities high enough to produce an $\HII$ region visible in SIGNALS, while the clusters themselves remain below the LEGUS detection threshold. \textsc{slug} creates samples of stellar populations using a Monte Carlo technique that takes into account the effect of stochasticity in the IMF, which is crucial for understanding stellar populations in the low-mass regime. The parameters we use to create the library of synthetic clusters are a slight modification of those used for the fiducial library described in \citet{2015ApJ...812..147K}: we use Padova stellar evolution tracks including TP-AGB stars \citep{2005ApJ...621..695V}, \textsc{starburst99} stellar atmosphere models \citep{1999ApJS..123....3L, 2005ApJ...621..695V}, a Milky Way extinction curve \citep{1989ApJ...345..245C}, a Kroupa IMF \citep{2001MNRAS.322..231K}, Solar metallicity, and 73\% ionising photons converted to nebular emission, with the remaining 27\% assumed to be absorbed by dust grains, following \citet{1997ApJ...476..144M}. The ages of the synthesised clusters are randomly drawn from a flat probability density function in age from $10^5$ to $10^7$ yr, the masses from a $dN/dM \propto M^{-2}$ distribution from 20 to $10^7$ $\Mo$, and the visual extinctions from flat distribution from $A_V=0-3$ \citep[see][and references therein for justifications of these choices of distribution]{2015ApJ...812..147K}.

Once the library is constructed, the next step is to determine if each synthetic star cluster would be classified as a cluster by LEGUS, and whether it would generate an $\HII$ region detectable by SIGNALS. For LEGUS, we use the selection criteria described by \citet{2015ApJ...815...93G} and summarised in \autoref{sec: starclusters}, assuming that a star cluster would not be detected in a given band if it has an apparent magnitude fainter than the 90\% completeness limit at the detection thresholds listed in Table 1 of \citet{2017ApJ...841..131A}; we use their stated limits for the centre pointing, as this includes the majority of detected cluster candidates. These detectable synthetic clusters are then considered classifiable by LEGUS if they have an absolute V band magnitude $ < -6$ mag. For SIGNALS, we treat a synthetic cluster as detectable if its predicted $\Ha$ luminosity, $\LHa$, exceeds the SIGNALS detection threshold for NGC~628 ($\log{\LHa}=35.65$). We estimate the $\Ha$ luminosity of the $\HII$ regions created by synthetic star clusters assuming
\begin{equation}
    \LHa = \alpha_{\rm H\alpha} f_g \QH
    \label{eq: 1}
\end{equation}
where $\alpha_{\rm H\alpha}= 1.37\times10^{-12}$ erg / photon is the conversion efficiency from ionising photons to H$\alpha$ emission expected for case B recombination \citep{1987MNRAS.224..801H, 1998ApJ...498..541K}, $\QH$ is the ionising luminosity of the synthetic star cluster, and $f_g=0.73$ is the fraction of ionising photons absorbed by gas rather than dust \citep{1997ApJ...476..144M}. We further reduce the $\Ha$ luminosity by applying a nebular extinction $A_{V,\rm neb}$ assuming that the ratio of nebular to stellar extinction $A_{V, \rm neb}/A_{V, \star}=2.27$ \citep{2000ApJ...533..682C}, where $A_{V,\star}$ is the stellar extinction drawn for that cluster.

\begin{figure}
    \centering
    \includegraphics[width = \columnwidth]{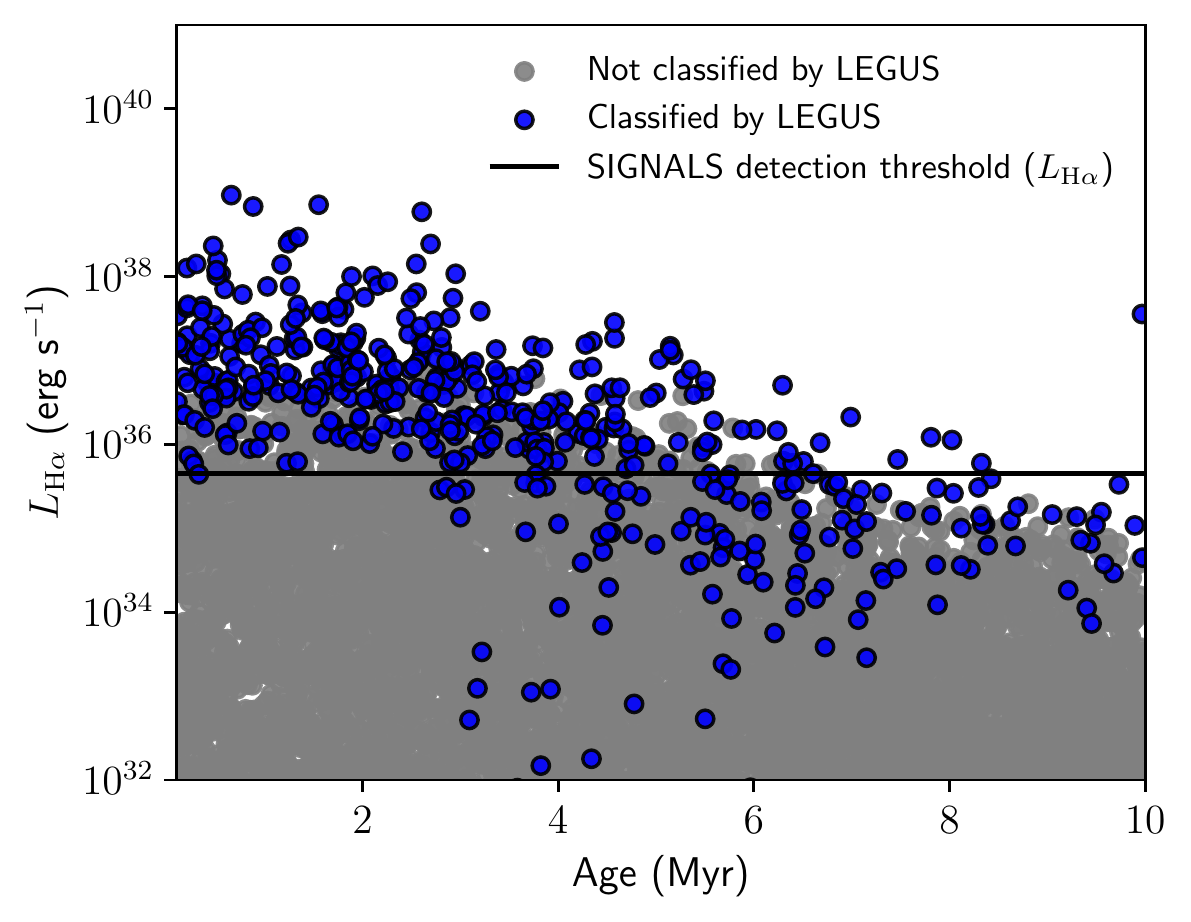}
    \caption{The theoretical $\Ha$ luminosity vs age distribution of $10^{5}$ synthesised YSCs from \textsc{slug}, showing star clusters that would be detected and classified in LEGUS (in blue), and star clusters that would not (in grey). Data presented in this plot are randomly chosen subsets of the complete analysis to minimise crowding. The horizontal dashed line in black shows the $\Ha$ detection limit of SIGNALS for NGC~628 ($\log{\LHa}=35.65$). We find 48\% orphan $\HII$ regions (i.e., grey data points above the detection limit of SIGNALS) in our synthetic clusters.}
    \label{fig: SLUG_completeness}
\end{figure}

\begin{table}
	\centering
	\begin{tabular}{ll} 
		\hline
		\hline
		Potential   & Potential contribution \\
		Explanation & to orphan $\HII$ regions  \\
		\hline
		Unclassified star clusters &  160 orphans \\
		Class~4 sources & 36 orphans \\
		Over-segmentation of $\HII$ regions  & $<20$ orphans \\
		Astrometric uncertainties & $<5$ orphans\\
        \hline
        Total (in theory)& $<221$ orphans\\
        Total (observed) & 165 orphans\\
        \hline
	\end{tabular}
	\caption{An overview of the analysis on possible causes of orphan $\HII$ regions, ordered from highest potential contribution to lowest. Details and justifications for their numbers are available in their respective sections. We note that the results derived in this section should \textit{not} be taken at face value as the representation of the true underlying processes that are creating these orphan $\HII$ regions.}
	\label{tab: orphan_overview}
\end{table}

\autoref{fig: SLUG_completeness} shows the age and theoretical H$\alpha$ luminosity distribution of $10^{5}$ synthesised star clusters. The expected fraction of orphan $\HII$ regions is calculated as the ratio of synthetic star clusters that are not detectable or classified by LEGUS but still produce detectable $\HII$ regions (grey points above the black horizontal line) to the total number of detectable $\HII$ regions (all points above the horizontal line). We find 48\% of the $\HII$ regions detectable by SIGNALS in our synthetic library are powered by star clusters that would not be classified by LEGUS. For the observed sample size of 334 observed $\HII$ regions, this translates to a theoretical number of $\sim$ 160 orphan $\HII$ regions. In comparison, we observe 165 orphan $\HII$ regions in the combined catalogue. We therefore conclude that the largest factor contributing to orphan $\HII$ regions are regions with ionising star clusters that were not classified by LEGUS.

\subsection{OB associations}
\label{sec: diffuse}
Finally, we discuss the possibility that the orphan $\HII$ regions are ionised not by compact star clusters or by single stars, but are instead powered by loose, gravitationally unbound stellar complexes known as OB associations \citep[e.g., see][]{1970AJ.....75..171L, 1997ApJ...476..144M}. These OB associations span tens to hundreds of parsecs \citep[][]{1995AstL...21...10M, 2004A&A...413..889B} and have low spatial densities ($\lesssim 0.1$ M$_\odot$ pc$^{-3}$), but nonetheless may be surrounded by dense gas and thus may create detectable $\HII$ regions \citep[see][]{1988AJ.....96.1874C}. Most of the OB associations themselves, however, would not be included in the LEGUS catalogue, as LEGUS is optimised to extract only compact sources \citep[see][]{2015AJ....149...51C, 2015ApJ...815...93G, 2017ApJ...841..131A}. Indeed, \citet{2017ApJ...841...92R} and \citet{2021MNRAS.508.5935B} studied the radius distribution of LEGUS clusters in NGC~628 and showed that LEGUS did not contain sources with radii more extended than $\sim10$ pc. These diffuse ionising sources may contribute not only to the orphan regions, but may also help explain $\HII$ regions that are populated by star clusters with seemingly underestimated ionising luminosity (see \aref{ap: peculiar}). 

We can estimate the potential contribution of ionising photons from these missed OB associations by constructing an ionisation budget for the entire FoV and checking whether the region as a whole satisfies balance between the injection of ionising photons and the output of $\Ha$ photons. To this end, we include all $\HII$ regions (symmetrical, asymmetrical, diffuse and transient), and LEGUS sources (Class 1, 2, 3, and Class 4 sources found in $\HII$ regions; see \autoref{sec: class4} and \autoref{sec: cluster_dist} for justifications). For star clusters, we compute their ionising luminosity $\QH$ using \texttt{cluster\_slug} (see \autoref{sec: cluster_dist} for more details); for $\HII$ regions, we estimate $\QHa$ using the equation given in \autoref{tab: terminology}. Overall, we find $\f \sim 0.98^{+0.02}_{-0.10}$, where $\f = 1 - \QHa/\QH$. A value of $\f > 0$ indicates that the total ionising flux required to create $\HII$ regions is sufficiently supplied from the photons produced from ionising sources. Thus, while diffuse ionising sources missed by LEGUS can potentially explain some orphans, they cannot dominate the total ionisation budget of the galaxy.


\subsection{Summary of catalogue completeness analysis}
\label{sec: completeness_summary}

In summary, we find that astrometric uncertainties in the catalogues, over-segmentation of $\HII$ regions, isolated massive stars, and OB associations will unlikely contribute substantially to the orphan $\HII$ regions (see \autoref{tab: orphan_overview} for an overview). In addition, the possibility of orphan regions arising from supernovae remnants, planetary nebulae, and DIG regions -- if present -- represents a minor contribution and will not drive the results we see, as we have discussed in \autoref{sec: hii}. Instead, we conclude that most orphans are $\HII$ regions whose star cluster companions are either not bright enough to be visually inspected in LEGUS, or that they were inspected but were then misclassified as non-clusters (Class 4). The analysis in this section shows that the orphan $\HII$ regions are not due to mistaken associations.
    
\section{The Lyman-continuum escape fraction}
\label{sec:escape_fraction}

In this section, we evaluate the LyC escape fraction $\f$ for $\HII$ regions that do have matching LEGUS sources, using two complementary approaches; one where we analyse each $\HII$ region individually (\autoref{sec: individual}), and one where we analyse the collective properties of the cluster population as a whole (\autoref{sec: population}).

\subsection{Individual $\HII$ region analysis}
\label{sec: individual}

\subsubsection{Cluster $\QH$ distributions}
\label{sec: cluster_dist}

The simplest approach to study $\f$ is to directly compare the ionising luminosity of star clusters to the ionising luminosity required to power the $\Ha$ emission of associated $\HII$ regions. We therefore begin by computing $\QH$ for star clusters in populated $\HII$ regions. To account for stochasticity in the computation of cluster properties, we use \texttt{cluster\_slug}, a module that is part of the \textsc{slug} software package. \texttt{cluster\_slug} uses a Bayesian method to compute the posterior probability density function (PDF) of age, mass, and extinction of star clusters, by taking the photometry of a set of data as input and comparing them to a library of synthesised star clusters. Then, we use the modified version of \texttt{cluster\_slug} described in \citet{2022A&A...666A..29D, 2021A&A...650A.103D} to estimate the cluster ionising luminosities. 

For the library of synthetic clusters, we use the same physical parameters as described in \autoref{sec: unclassified} (i.e., atmosphere models, extinction curve, IMF, and metallicity), with the exception of adopting the Geneva evolutionary tracks \citep{2012A&A...537A.146E, 2012A&A...542A..29G}. Both Padova and Geneva models provide a relatively good agreement for models of stars close to the main-sequence, with the Padova models providing superior treatment for stars with ages $> 10$ Myr, where the contribution of red supergiants and asymptotic giant branch stars become important \citep[e.g., see][]{2005AIPC..761...39L}. For the purpose of our study, however, since the majority of targeted clusters are younger than $\sim 10$ Myr, we use the Geneva tracks, which are more tuned to reproduce observations of young, massive stars. However, we also present results derived using Padova tracks in \aref{ap: alternative_libraries}, and show that they lead to similar qualitative conclusions.

We again assume a flat prior on the extinction from $A_V=0-3$. However, for the purposes of this analysis we adopt slightly different priors on mass and age than \citet{2015ApJ...812..147K}: we take as priors $dN/dM\propto M^{-1}$ in mass from 20 to $10^7 \Mo$, $dN/dT \propto T^0$ (i.e.~flat) for $T < 10^{6.5}$ yr, and $dN/dT\sim T^{-1}$ for $T > 10^{6.5}$ yr. The mass prior is shallower and the age prior steeper than \citeauthor{2015ApJ...812..147K} used for their analysis of the full LEGUS catalogue because their data set includes all clusters in the galaxy, and therefore contains a larger fraction of older clusters; by contrast, we are selecting not all clusters, but only those associated with bright $\HII$ regions, which motivates us to choose priors somewhat more weighted to younger, more massive clusters. Indeed, we compared between our choice of priors and those implemented in \citet{2015ApJ...812..147K}, and find that ours can, in general, provide estimations of cluster ionising luminosity with smaller uncertainty, thereby allowing more data points for further analysis (see \autoref{fig: LHa_QH0_wCons}). Finally, for each star cluster we use \texttt{cluster\_slug} to evaluate the median (50th percentile) and 1-$\sigma$ uncertainty (16th -- 84th percentile) for its ionising luminosity, $\QH$, given the measured photometry. We compare the output to the $\Ha$ luminosity of $\HII$ regions with which these clusters are associated.

\begin{figure}
    \centering
    \includegraphics[width = \columnwidth]{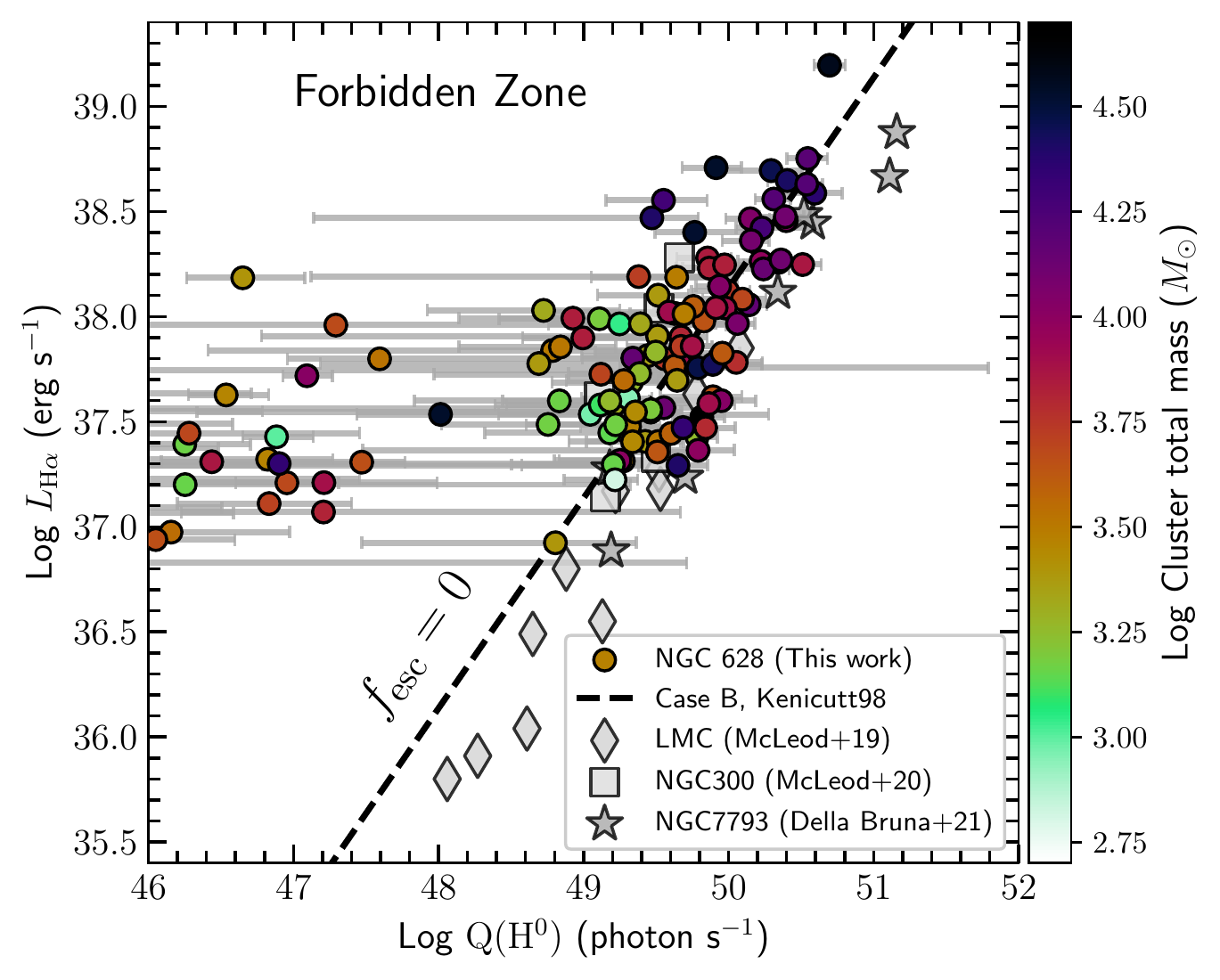}
    \caption{The observed $\Ha$ luminosity of $\HII$ regions ($\LHa$) versus the predicted photon flux of their ionising sources, $\QH$, before removing sources with large uncertainties (see \autoref{sec: individual}). Coloured circles are $\HII$ regions in NGC 628 (this study) where the photon flux of corresponding star clusters is computed with modified \texttt{cluster\_slug}. The colourbar indicates the sum of the 50th percentile estimates for the masses of star clusters associated with the corresponding $\HII$ regions. The $\QH$ values shown by the points correspond to the median of the PDF, with horizontal errorbars indicating the 15.9th -- 84.1th percentile ranges. The errorbars are only displayed for a subset of the data to minimise crowding. Diamond, square, and star symbols without error bars are literature values for $\HII$ regions in the LMC \citep{2019MNRAS.486.5263M}, NGC 300 \citep{2020ApJ...891...25M}, and NGC 7793 \citep{2021A&A...650A.103D} respectively. The black dashed line indicates the luminosity-photon flux conversion expected for Case B recombination, i.e., where $\f = 0$ \citep[see][]{1987MNRAS.224..801H, 1998ApJ...498..541K}. The plot shown here is contaminated by various sources of uncertainty; we remove them before further analysis (see discussion in \autoref{sec: individual}).}
    \label{fig: LHa_QH0_noCons}
\end{figure}

\begin{figure}
    \centering
    \includegraphics[width = \columnwidth]{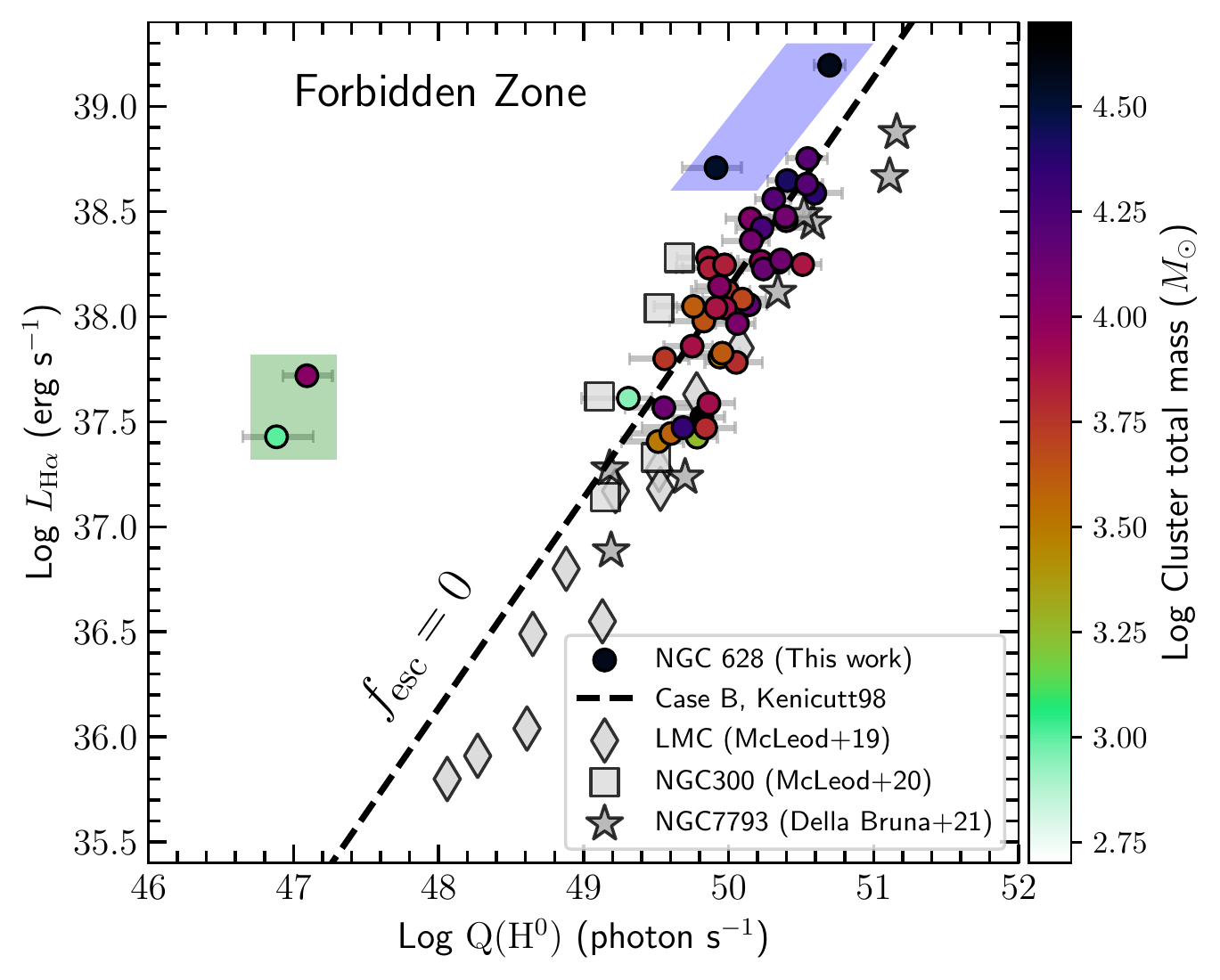}
    \caption{Same as \autoref{fig: LHa_QH0_noCons}, but only retaining $\HII$ regions for which the 68\% confidence interval on the ionising luminosity of the associated star clusters is $<0.5$ dex (see \autoref{sec: individual}). The green and blue polygons, coloured differently to indicate different origins of peculiarity (see discussion in the main text, and in \aref{ap: peculiar}), highlight data points which will not be included in our calculations of $\f$. The remaining data points are in agreement with theoretical expectations (i.e., whose $\QH$ distribution lie on or below the dashed line).}
    \label{fig: LHa_QH0_wCons}
\end{figure}

Before diving into the analysis, we make four additional constraints regarding the sample of populated $\HII$ regions. First, while observations indicate that essentially all clusters older than 10 Myr have cleared their environments \citep[e.g.,][]{2011ApJ...729...78W, 2015MNRAS.449.1106H,2018MNRAS.481.1016G,2019MNRAS.490.4648H, 2021ApJ...909..121M, 2021MNRAS.504..487K, 2022MNRAS.512.1294H}, and that the typical timescale for gas clearing due to feedback from massive stars in clusters is $\lesssim$ 5 Myr \citep[e.g.,][]{2003ApJ...596L.135B,2013MNRAS.430..234D,2016ApJ...821...38S, 2022MNRAS.509..272C, 2022MNRAS.516.3006K}, we do not impose as a prior that the age must be smaller than some value. The reason for this is that we expect at least some of our cluster-$\HII$ region associations to be chance alignments along the line of sight, or even clusters that are physically located within $\HII$ regions but are so located by chance, not because they are responsible for creating that $\HII$ region. We wish to allow our fits to determine older ages in such cases, and these cases are, as we see below, immediately apparent because they lie far from the physically plausible part of parameter space. Second, in this portion of our analysis, we do include Class~4 sources present within $\HII$ regions, for the reasons discussed in \autoref{sec: class4}: at least some Class~4 sources are true clusters, and contribute ionising fluxes to their associated $\HII$ regions. We err on the side of inclusion rather than exclusion because almost all potential interlopers that contribute to the Class~4 population (e.g., background galaxies) are relatively faint in the UV, and this translates to our fitting method assigning them quite low ionising luminosities. As with older clusters, the extent that these are false associations becomes obvious when we compare them to the observed $\Ha$ luminosity. We also present results from an analysis excluding Class~4 clusters in \aref{ap: alternative_libraries}, and show that doing so does not yield qualitatively different results. Third, we discard data points where a star cluster is potentially associated with multiple $\HII$ regions. Such multiple associations are possible because, while the SIGNALS $\HII$ region decomposition algorithm assigns each pixel of the image to a unique $\HII$ region, the SIGNALS catalogue approximates irregular-shaped $\HII$ regions as circular regions, and there is no guarantee that these circles do not overlap. Overlapping regions that also contain a LEGUS cluster are therefore excluded, since we cannot make unique cluster-$\HII$ region associations for them. Finally, we exclude cluster-$\HII$ region associations whose distance from the centre of the $\HII$ region to the nearest FoV boundary is $\leq 1.5 R$, where $R$ is the radius of $\HII$ region. We impose a generous threshold here to avoid the inclusion of $\HII$ regions whose ionising sources are potentially not observed by the LEGUS survey. These constraints further reduce the number of $\HII$ region candidates from 169 to 139.

We present the initial result in \autoref{fig: LHa_QH0_noCons}, showing the $\Ha$ luminosity of all 139 $\HII$ region versus the ionising luminosities of its associated sources; where more than one cluster falls within an $\HII$ region, we sum their ionising luminosities. This plot can be understood as follows: if every LyC photon is absorbed by hydrogen atoms in the envelope of an $\HII$ region, and the $\HII$ region reaches ionisation-recombination balance, then all $\HII$ regions should lie on the \citet{1998ApJ...498..541K} line. $\HII$ regions that lie below this line have lower $\Ha$ luminosities than expected from the LyC flux of their source stellar population. This indicates that ionising photons are not completely absorbed by gas in the region. Conversely, $\HII$ regions that lie above the line substantially should be forbidden, as the inferred ionising luminosity is insufficient to produce the observed $\Ha$ luminosity.\footnote{Assuming no systemic errors, $\HII$ regions may still lie above the line due to ionisation sources other than star clusters.}

Two noticeable features can be inferred from our initial result in \autoref{fig: LHa_QH0_noCons}. First, we remark that most data points have large 1-$\sigma$ uncertainties in ionising luminosity. This is because \texttt{cluster\_slug}'s 68\% confidence range properly accounts for the uncertainties from IMF stochasticity and from degeneracies in age, mass and extinction due to the cluster's location in colour space \citep[see][]{2015ApJ...812..147K}. Moreover, when the PDF is multi-peaked, which is a relatively common outcome of a full stochastic analysis \citep[e.g., see][and \aref{ap: visualise_QH0}]{2012ApJ...750...60F, 2015MNRAS.452.1447K}, the 16th to 84th percentile range becomes very broad, since it must expand to encompass both peaks. Additionally, unless the star clusters are massive enough to fully sample the IMF, the ionising luminosity diagnosed from optical/UV photometry will come with large uncertainty. This is due to the photometry of low-mass clusters being dominated by a few massive stars, and colours of individual stars in optical/NUV bands are very insensitive to mass once the stellar effective temperature is high enough to shift its spectral energy distribution peak out of the observed bands. A 25 $\Mo$ star has nearly the same colour in available broadband optical/UV bands as a 75 $\Mo$ star, but the 25 $\Mo$ star will have an ionising luminosity many orders of magnitude lower than the 75 $\Mo$ star.

Second, we find that data points where the entire 1-$\sigma$ confidence interval lies above the \citet{1998ApJ...498..541K} line are mostly clusters that are at the low mass end of the observed mass distribution, and with ages $t > 5$~Myr where the inferred ionising luminosity has large variance. This result is expected and can be interpreted in two ways: one implication is that these clusters have multi-peaked or skewed PDFs, due to the stochasticity of IMF sampling in the low-mass regime. Thus, although the median of the ionising luminosity lies to the left of the \citet{1998ApJ...498..541K} line, there is nonetheless a reasonable chance that the true ionising luminosity lies to the right of the line. Another possible explanation is that these are not real associations, i.e., the detected star clusters are chance alignments with their $\HII$ region pairs. This is likely to be the case for most of the data points that are located far above ($\gtrsim 0.5$ dex) the \citet{1998ApJ...498..541K} line. In these cases, it is likely that the star clusters that are responsible for the ionisation of these $\HII$ regions are unclassified because, while bright at ionising wavelengths, they fall below the LEGUS catalogue's detection limits in the optical bands, while another cluster that is detectable by LEGUS (i.e., that is bright in optical but faint in ionising photons) is associated with the $\HII$ region due to chance alignment. 

To mitigate this uncertainty, we truncate the sample by retaining only $\HII$ regions for which the 1-$\sigma$ uncertainty on the ionising luminosities of the associated star clusters is $< 0.5$ dex. We show this reduced sample in \autoref{fig: LHa_QH0_wCons}. We note that, by imposing this error constraint we are effectively retaining the young ($\lesssim 5$ Myr) and high-mass end ($ \gtrsim 10^{4} \Mo$) of the sample, which are the clusters with the least stochastic variation. Even with this cut, some data points remain well within the forbidden zone (highlighted with green and blue polygons in \autoref{fig: LHa_QH0_wCons}); we discard these as well, and defer a discussion of them to \aref{ap: peculiar}. This reduces the number of data points from 139 to 42. These represent a sample of $\HII$ regions for which we can, based solely on the photometry, estimate the ionising luminosity of the associated star clusters with relatively high confidence.

\subsubsection{Probability distribution and caveats for $\f$}
\label{sec: f_prob_dist}

For our remaining sample, we evaluate the LyC escape fraction by drawing $10^6$ Monte Carlo samples from the converged Monte Carlo Markov Chains computed by \texttt{cluster\_slug} to represent the probability distribution for $\QH$; where multiple clusters fall within a single $\HII$ region, we draw samples for each cluster and sum to produce $10^5$ realisations for the \textit{total} ionising luminosity. For each realisation, we compute the escape fraction as
\begin{equation}
    \f = 1 - \frac{\LHa}{\alpha_{\rm H\alpha} \QH},
    \label{eq:fesc}
\end{equation}
where $\QH$ is the ionising luminosity for that sample, $\LHa$ is the observed H$\alpha$ luminosity, and $\alpha_{\rm H}$ is the conversion from ionising photons to H$\alpha$ luminosity; intuitively, this equation simply measures how far below the dashed line in \autoref{fig:  LHa_QH0_wCons} a given data point lies. The result is a PDF of $\f$ for each cluster.

We pause here to point out two features of $\f$. The first is that our definition of $\f$ includes any mechanism that causes a given ionising photon not to ionise gas within the $\HII$ region. While this could mean that the photon physically escapes the $\HII$ region, it could also mean that the photon was absorbed by a dust grain within the $\HII$ region; our analysis is not capable of distinguishing between these two possibilities. Second, there is nothing in \autoref{eq:fesc} that enforces $\f \geq 0$. Negative values of $\f$ correspond to estimates of the ionising luminosity $\QH$ small enough that there would be insufficient photons to power the observed $\Ha$ emission even for $\f = 0$. Given the rather broad PDFs of $\QH$ that we obtain from photometry, it is inevitable that in some cases parts of the PDF will extend into the $\f<0$ region. We could avoid this by imposing as a prior that $\f\geq 0$, but doing so would artificially suppress the width of our confidence intervals, and might conceal problems in our analysis. For this reason, we choose to adopt a purely flat prior of $\f$, allowing our analysis to produce $\f<0$ even though we know such values are unphysical. However, for completeness we also repeat our analysis with an $\f \geq 0$ prior in \aref{ap: alternative_libraries}, and show there that our qualitative conclusions are similar to those derived in the main paper.

\begin{figure}
    \centering
    \includegraphics[width = \columnwidth]{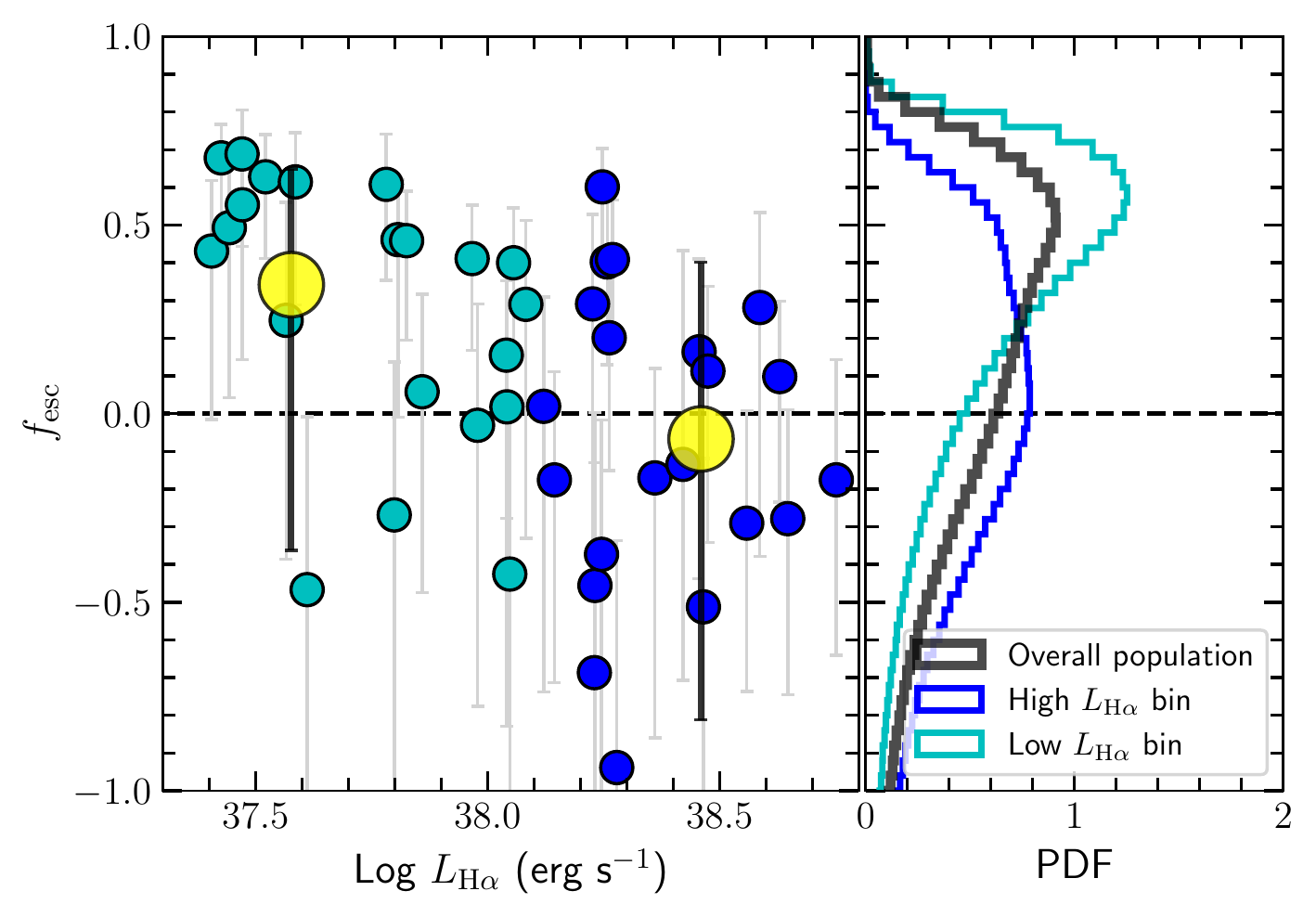}
    \caption{The LyC escape fraction, $\f$, of individual $\HII$ regions after truncating the sample (see \autoref{fig: LHa_QH0_wCons}), using the technique describeed in \autoref{sec: cluster_dist}. In the left panel, circles correspond to the median of the PDF of $\f$, coloured in cyan and blue to indicate low and high $\LHa$ bins, with grey errorbars indicating the 68\% confidence interval. The dashed line indicates $\f = 0$. Negative values of $\f$ correspond to estimates of the ionising luminosity $\QH$ small enough that there would be insufficient photons to power the observed $\Ha$ emission even for $\f =0$. For the histogram plot, the combined PDF is shown for each bin, along with an overall PDF of $\f$ (in black) across the truncated sample. Yellow data points illustrate median values of $\f$ for low- and high-$\LHa$ bins, where $\f = 0.34^{+0.31}_{-0.70}$ and $-0.07^{+0.47}_{-0.75}$ respectively.
   }
    \label{fig: Fesc_PDF}
\end{figure}

In order to study the possible relationship between $\f$ and $\LHa$, we divide the samples into low and high $\LHa$ bins, defined by the intervals $37.41 \leq \log L_{\rm {H\alpha}_{,low}} ({\rm erg~s}^{-1})\leq 38.06$ and $38.06 \leq \log L_{\rm {H\alpha}_{,high}} \leq 38.75$, respectively. These bins are chosen so that each contains roughly half of the sample. We present both the distribution of $\f$ for individual $\HII$ regions and the combined PDF of $\f$ for each luminosity bin in \autoref{fig: Fesc_PDF}. Our analysis yields an overall escape fraction of $\f = 0.13^{+0.43}_{-0.76}$, with  $\f = 0.34^{+0.31}_{-0.70}$ and $-0.07^{+0.47}_{-0.75}$ for low- and high-$\LHa$ bins respectively; the figures we quote here are the 50th percentile values, with error estimates corresponding to the 68\% confidence range. 

\subsection{Population analysis}
\label{sec: population}

The preceding section attempts to derive an escape fraction separately for each $\HII$ region, and then constructs a mean escape fraction for the entire population of $\HII$ region simply by summing those. However, given the rather broad PDFs of ionising luminosity that our photometric analysis returns for each individual cluster, it may be more reliable to attempt to constrain the escape fraction of the population as a whole directly. To this end, let us assume that there exists an overall escape fraction $\f$ across the $\HII$ region population; our goal is to determine the probability density function of this overall $\f$ given the observed $\LHa$ and photometry.

Our basic approach comes from Bayes's theorem: the posterior probability density for the escape fraction, $\f$, is proportional to the product of the prior probability and the likelihood function, which gives the probability of the data given the model. To determine the likelihood, first consider the simplest case of a single $\HII$ region with observed $\Ha$ luminosity $\LHa$ that is associated with a group of clusters, characterised by a set of photometric measurements $\{\fphot\}$. From our \texttt{cluster\_slug} analysis, for each $\HII$ region we can compute a PDF of ionising luminosity $p_{\textsc{slug}} \br{\QH|\{\fphot\}}$ for the associated star clusters from the photometry exactly as in \autoref{sec: individual}. If the true escape fraction is $\f$, then the corresponding PDF of the expected $\Ha$ luminosity has the same shape, with each ionising luminosity $\QH$ simply mapped to a corresponding $\Ha$ luminosity $\LHa = (1 - \f) \alpha_{\rm H\alpha} \QH$. Mathematically, this can be  written as  
\begin{align}
    p\br{\LHa \mid \f, \{\fphot\}} \propto  p_{\textsc{slug}}  \left.\br{\QH|\{\fphot\}} \right|_{\QH = \frac{\QHa}{1-\f}}
\end{align}
where for convenience we have defined $\QHa = \LHa/\alpha_{\rm H\alpha}$ as the ionising luminosity required to drive a given $\Ha$ luminosity in the limit of zero escape fraction. This gives the likelihood function for a single $\HII$ region. The generalisation to a population of $\HII$ regions is straightforward, since each one is independent: given a set of $N$ measured $\HII$ region $\Ha$ luminosities $\LHa_{,i}$, where $i = 1\ldots N$, and a corresponding set of photometric measurements $\{\fphot\}_i$ for the clusters associated with those $\HII$ regions, the likelihood function is
\begin{multline}
    p\br{\{\LHa\} \mid \f, \{\{\fphot\}\}}  \\
    \propto \displaystyle \prod^{N}_{i=1} p_{\textsc{slug}} \left.\br{\LHa_{,i}|\{\fphot\}_i} \right|_{\QH = \frac{\QHa_i}{1-\f}}
\end{multline}
where $\{\LHa\}$ is the set of all observed $\Ha$ luminosities and $\{\{\fphot\}\}$ is the set of all observed photometric magnitudes for all clusters in all $\HII$ regions; note that the double curly braces on $\{\{\fphot\}\}$ indicate a nested list -- all clusters observed in all filters.

\begin{figure}
    \centering
    \includegraphics[width = \columnwidth]{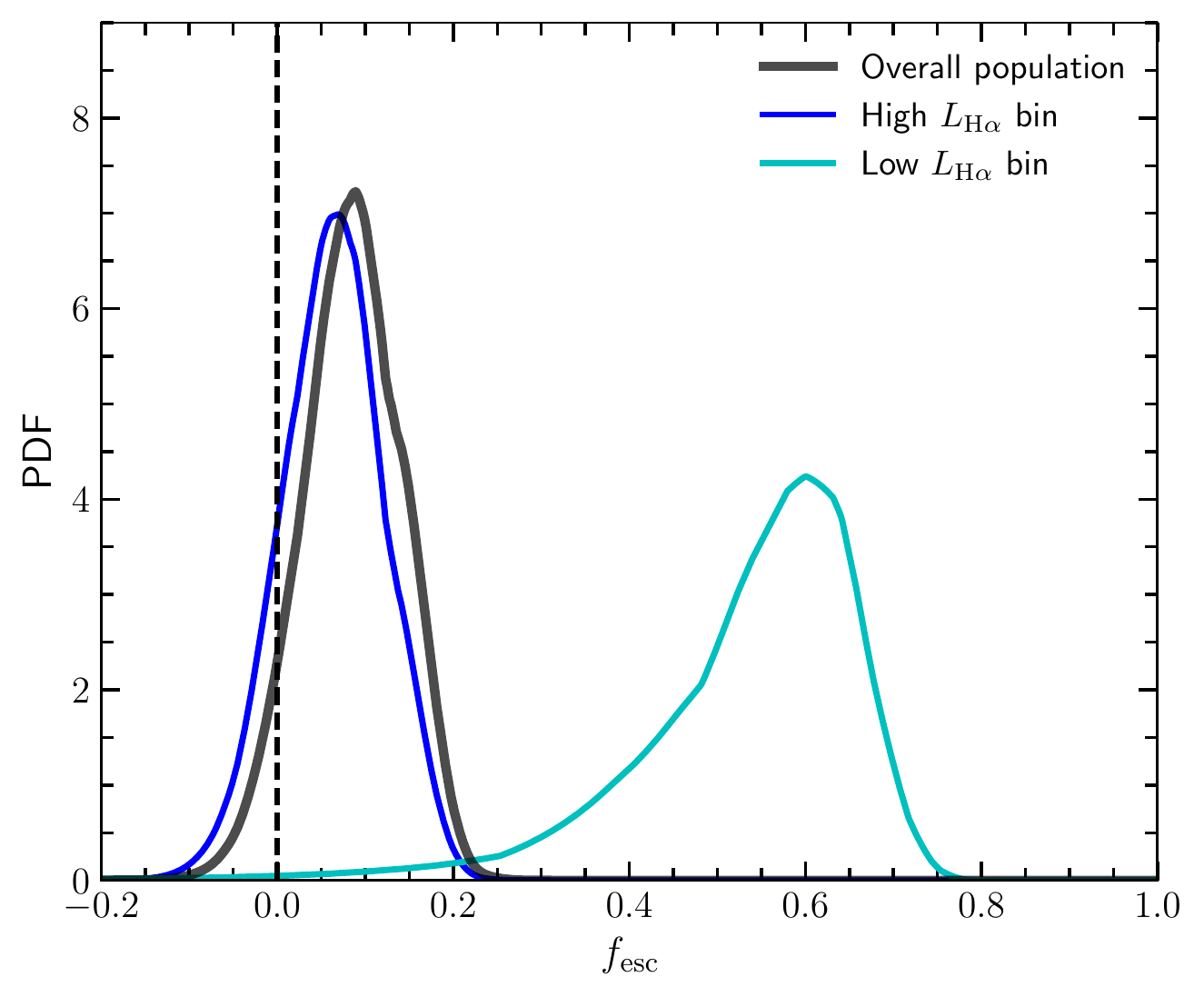}
    \caption{Posterior PDF of $\f$ derived using the technique described in \autoref{sec: population}.  Histograms coloured in cyan and blue corresponds to low- and high-$\LHa$ bins. The dashed line indicates $\f = 0$. The PDF for the whole population is shown in black. Similar to \autoref{fig: Fesc_PDF}, we find that the overall escape fraction across the population of low-$\LHa$ $\HII$ regions ($\f = 0.56^{+0.08}_{-0.14}$) is higher than that of the population of high-$\LHa$ regions ($\f = 0.06^{+0.06}_{-0.06}$).}
    \label{fig: Fesc_Bayesian}
\end{figure}

For simplicity, we adopt a flat prior on $\f$, which means that the posterior PDF for the escape fraction is simply proportional to the likelihood function. Thus our final expression for the posterior PDF of the escape fraction is
\begin{multline}
    p\br{\f \mid \{\LHa\}, \{\{\fphot\}\}}  \\
    \propto \displaystyle \prod^{N}_{i=1} p_{\textsc{slug}} \left.\br{\LHa_{,i}|\{\fphot\}_i} \right|_{\QH = \frac{\QHa_i}{1-\f}}
\end{multline}
We can evaluate this expression either for the entire population of $\HII$ regions, or for subsets thereof binned by $\LHa$ or in any other way. We plot our posterior PDFs of $\f$ for both the full sample and for the low- and high-$\LHa$ sub-samples in \autoref{fig: Fesc_Bayesian}. We find an overall escape fraction of $\f = 0.09^{+0.06}_{-0.06}$ for the full sample, with $\f = 0.56^{+0.08}_{-0.14}$ and $0.06^{+0.06}_{-0.06}$ for the low- and high-$\LHa$ bins respectively; as in \autoref{sec: individual}, the numbers we quote are the 50th percentile value with error bars giving the 68\% confidence range. We further discuss the implication of this result in \autoref{sec: linking}.

\section{Discussion}
\label{sec:discussion}
\subsection{Improving the number of observed populated H~\textsc{ii} regions}
\label{sec: vmag_cut}

\begin{figure}
    \centering
    \includegraphics[width = \columnwidth]{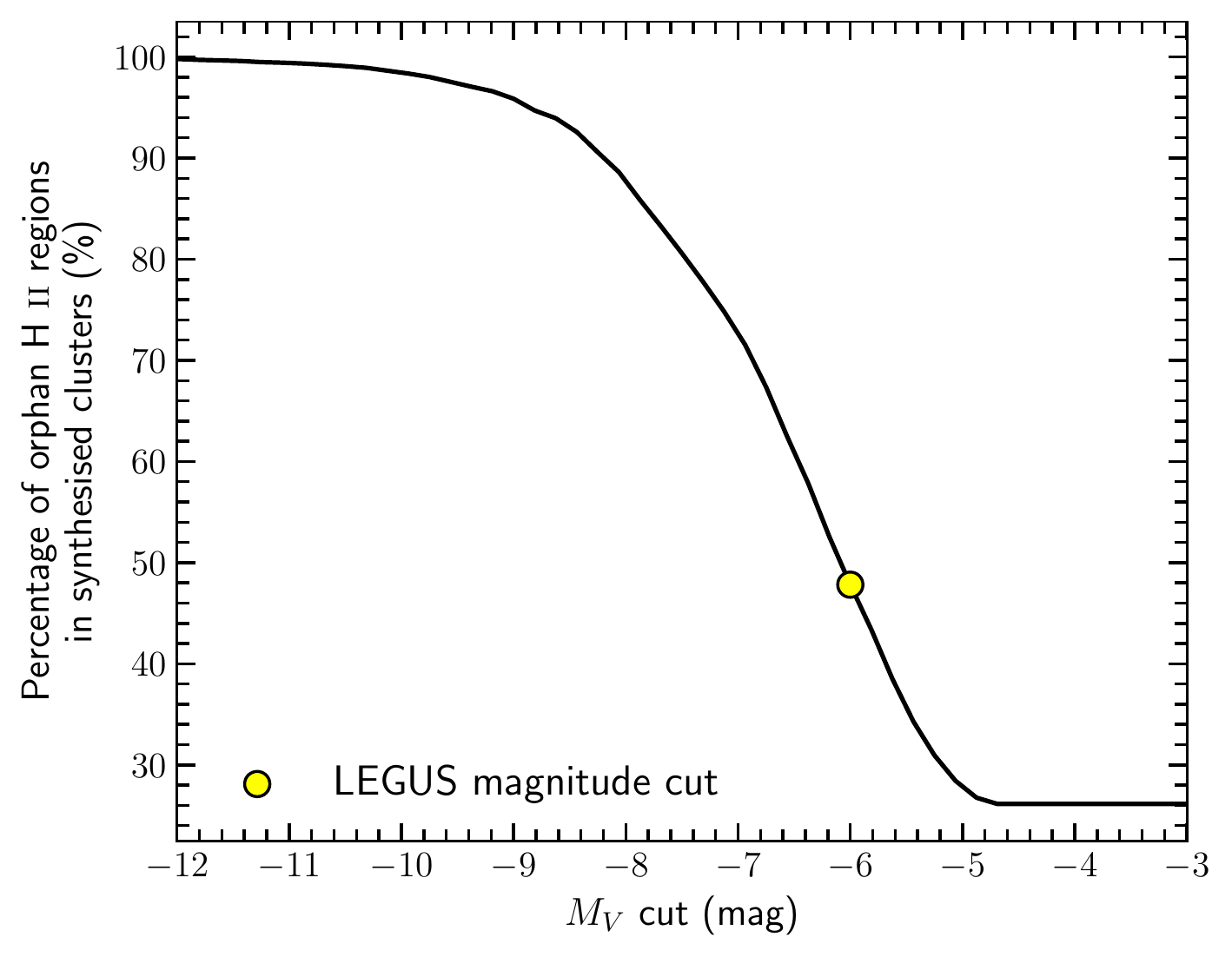}
    \caption{The percentage of orphan $\HII$ regions as a function of V-band absolute magnitude cut in the synthesised YSCs using \textsc{slug}. Yellow dot indicates the visual inspection magnitude cut in LEGUS. The predicted number of orphan $\HII$ regions decreases as $M_V$ cut increases; this is because we are allowing visual inspection on fainter clusters.}
    \label{fig: vmag_cut}
\end{figure}

In \autoref{sec:completeness} we show that the orphan $\HII$ regions are mainly a consequence of $\HII$ regions whose associated star clusters are either less luminous than the LEGUS visual inspection threshold ($M_V<-6$), or that they were inspected and classified as non-clusters (Class 4). Before moving on, it is thus necessary to provide a resolution to this problem if we wish to harness the full potential of oncoming LEGUS-SIGNALS observations. What can we do to improve the number of observed populated $\HII$ regions?

There are two possible ways around this issue. The first of these is to allow visual inspection of fainter ($M_V\geq -6$) LEGUS sources, which in turn can be approached both via observational and theoretical grounds. On the observational side, we begin by analysing the location of LEGUS sources that do not meet the magnitude threshold required for visual inspection (Class 0; \citealp[see][]{2015ApJ...815...93G}). We overlay these sources onto a map of $\HII$ regions that do not contain any of Class 1, 2, 3 or 4 LEGUS sources, and find that 13 out of 129 ($\sim 10\%$) of these $\HII$ regions contain Class 0 sources. While we note that Class 0 sources may include spurious detections, this nevertheless suggests that they may contain true clusters.

On the other hand, we can make theoretical predictions of how inspecting fainter clusters would decrease the number of orphan $\HII$ regions by repeating the completeness analysis in \autoref{sec: unclassified}, with a simple change in the cluster selection criteria: instead of using the LEGUS magnitude cut, $M_V=-6$, we experiment with magnitude cuts ranging from $M_V = -12$ to $M_V = -3$. For each $M_V$, we calculate the percentage of orphan $\HII$ region in the synthesised star clusters and show the result in \autoref{fig: vmag_cut}. We find that changing the magnitude cut from $M_V = -6$ to $M_V = -5$ will result in a decrease in the percentage of orphan $\HII$ regions (from 48\% to 28\%) in the synthesised star clusters. Further allowing clusters that are fainter than $M_V = -4$ seems to have no effect on the orphan $\HII$ regions: we suspect that these star clusters are likely unclassified due to the completeness limits in other photometric filters, as $M_V = -6$ is not the only cluster selection criterion. Cluster candidates require detection in at least two filters (the V band and either B or I band), and must be detected in 4 bands total, implying that the cluster must be detected either in the U or the UV band. This means that clusters with low mass and moderate extinction could, in theory, be detectable in the B, V and I bands, but not in the bluer bands. Therefore, even if one inspects fainter clusters, some clusters  will be excluded from the LEGUS catalogue simply because they will not be detected at all.

Finally, it is worth noting that while varying the magnitude cut seems reasonable on paper, it is impractical. This is because the number of potential clusters that have to go through human visual classification is significantly higher, since the cluster luminosity function of NGC 628 scales as $dN/dL\approx L^{-\beta}$, where $\beta\approx 2$ \citep[see][]{2017ApJ...841..131A}. Thus, the time-consuming nature of this labour-intensive task limits the choice of the magnitude cut.\footnote{However, for those interested to study only clusters that are located in $\HII$ regions, it is still worthwhile to inspect all detected sources regardless of their magnitude to allow for more complete coverage of cluster candidates.} Instead, machine learning techniques could be implemented in the LEGUS pipeline to ensure reproducibility of cluster classifications and a significantly higher classification speed \citep[e.g., see][]{2019MNRAS.483.4707G, 2020MNRAS.493.3178W, 2021ApJ...907..100P, 2022ApJ...935..166L}.

This brings us to the second method. As demonstrated in \autoref{sec: class4}, we find, via statistical and visual techniques, that a fraction of Class~4 sources are likely true clusters after we combine and inspect spatial information of sources in the LEGUS catalogue with an $\Ha$ map from SIGNALS, and look for areas of coincidence. An important corollary of this result is that we can further improve the accuracy of visually-identified clusters (especially clusters $< 5$Myr) in future surveys by overlaying LEGUS sources on a resolved $\Ha$ image.

\subsection{Improving measurement of $\f$: ways forward}
\label{sec: ways_forward}

The paper thus far shows that \texttt{cluster\_slug} is a powerful tool to infer stellar properties (e.g., age, mass, $\QH$) from five-band photometry alone, but also suggests some weaknesses and possible ways forward. The first is the lack of far-ultraviolet (FUV; $\approx 8-13.6$ eV) data: FUV surveys of stellar populations provide much stronger constraints on the properties of young stellar populations than observations at longer wavelengths alone \citep[e.g., see][]{2008ApJ...683.1006K, 2011ApJ...727..100W, 2016MNRAS.457.4296W, 2016ApJ...823...38S, 2022MNRAS.510.4819S}. In our case, as discussed in \autoref{sec: cluster_dist}, the dominant contributor to our uncertainties on ionising luminosities is the fact that all massive stars have similar colours in the optical to NUV bands, meaning that our data offer limited constraints for clusters whose light is dominated by one or a few such stars. The addition of FUV observations would reduce this issue, greatly improving our ionising flux estimates and therefore our $\f$ measurements. Existing FUV data from \textit{GALEX} \citep[Galaxy Evolution Explorer;][]{2005ApJ...619L...1M} are not suitable for this purpose due to their limited spatial resolution, which makes it essentially impossible to assign \textit{GALEX}-detected FUV emission uniquely to particular LEGUS clusters. Instead, FUV data with HST-like resolution are required. One possible future source for such data is the \textit{UVEX} survey \citep[Ultraviolet Explorer;][]{2021arXiv211115608K}, which is designed to perform a multi-cadence synoptic all-sky survey $50-100$ times deeper and at significantly higher resolution than \textit{GALEX} in the FUV/NUV.

A second source of uncertainty in our analysis the loss of LyC photons to dust absorption, and this too could be improved with future data. The inclusion of high-spatial-resolution infrared observations such as from the PHANGS-\textit{JWST} programme \citep{2022arXiv221202667L} and the incoming \textit{JWST}-FEAST programme \citep{2021jwst.prop.1783A}, among several others, will allow us to trace dust emissions around clusters. On the other hand, comparison between our observations and stellar feedback codes such as \textsc{warpfield} \citep{2017MNRAS.470.4453R, 2019MNRAS.483.2547R}, which allows detailed study of  clusters and their feedback effect on surrounding natal clouds, will also provide useful insight on the dust content. This in turn will allow better estimates of the fraction of LyC photons that are absorbed by dust grains.

\subsection{Linking between $\f$ and local environments}
\label{sec: linking}

\subsubsection{Notes on the overall $\f$ and $\fDIG$}
\label{sec: fesc_implication}

In \autoref{sec: population} we report an overall LyC escape fraction of $\f = 0.09^{+0.06}_{-0.06}$ across our population of $\HII$ regions. The escape fraction is lower than what is usually reported for $\HII$ regions in nearby dwarf galaxies \citep[$0\lesssim \f \lesssim 0.51$;][]{2019MNRAS.486.5263M, 2020ApJ...891...25M, 2020ApJ...902...54C} and spiral galaxies ($\f\sim 0.67$; \citealt{2021A&A...650A.103D}; see also \citealt{2022A&A...666A..29D}). We note that these values of $\f$ should not be compared at face value as they represent a wide range of galactic environments, and that they were evaluated using different methods; however, it is still necessary to address any biases present in our analysis, and their potential effect on $\f$. First, in this study we discard $\HII$ regions that are deemed diffuse and transient regions by the SIGNALS survey. While this is important for the analysis since we want to minimise contamination from highly dispersed regions that are likely DIG-dominated \citep{2018MNRAS.477.4152R, 2019MNRAS.489.5530R}, it implies that the majority of our regions may be the ones that are compact and ionisation-bounded. Thus, one may expect that the overall LyC escape fraction is much closer to zero. There is also a secondary factor, which is that the selected $\HII$ regions are limited by the LEGUS FoV: we are essentially observing the inner half of the galaxy disc, thus more towards regions of higher metallicity, extinction and dust-gas mass ratio. This may also point towards a higher likelihood of having a lower overall $\f$.

It is also illuminating to compare $\f$ and the fraction of $\Ha$ emission in the DIG ($\fDIG$). Estimates of $\fDIG$ in NGC~628 range from $\approx 0.25 - 0.5$ \citep[see][]{2013ApJ...762...79C, 2020MNRAS.493.2872C, 2022A&A...659A..26B}, much larger than the overall $\f$ determined in this study. However, given the systematic uncertainties in previous studies we do not conclude that this is inconsistent with the idea that the DIG is powered by leaking $\HII$ regions; rather, we posit the possibility that the DIG includes a subdominant contribution from harder ionising photons from old stars \citep[similar to][]{2022A&A...659A..26B}. The reason is that we have only taken into account clusters associated with $\HII$ regions, which tend to be very young ($< 5$ Myr). We do not consider older clusters that are not associated anymore with compact $\HII$ regions, but nevertheless these can still contribute to the DIG.

\subsubsection{$\f$ and the properties of H~\textsc{ii} regions and clusters}
\label{sec: fesc_vs_prop}

\begin{figure}
    \centering
    \includegraphics[width = \columnwidth]{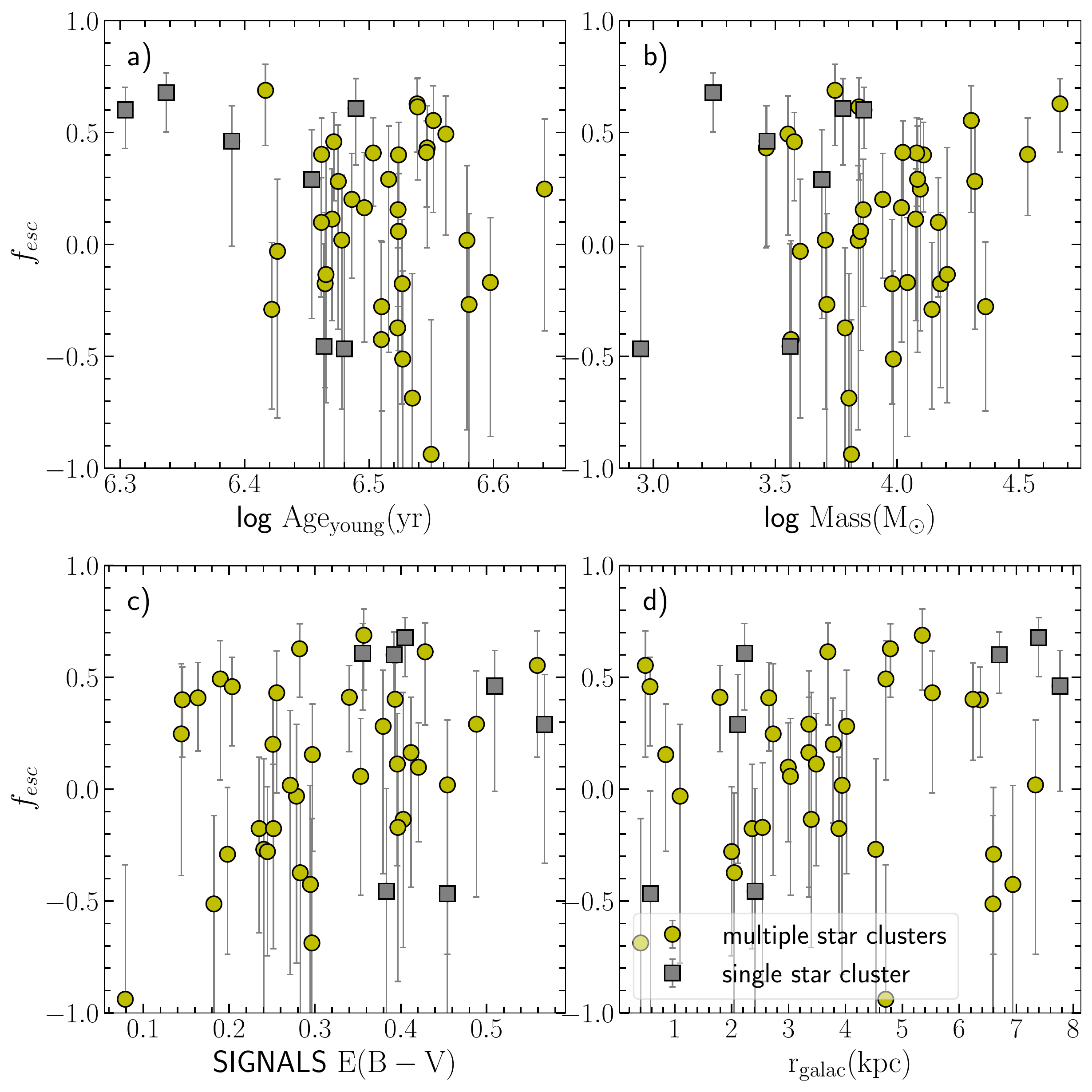}
    \caption{Properties of $\HII$ regions and their associated star clusters vs.~the LyC escape fraction, $\f$. We plot $\HII$ regions found to be associated with single clusters as grey squares, and those associated with multiple clusters as yellow circles. The errorbars represent 1-$\sigma$ uncertainties on $\f$; we retain distributions of $\f$ that extend into the $\f<0$ region for reasons discussed in \autoref{sec: f_prob_dist}. The top two panels show respectively the median age across clusters in the younger age bin (age$_{\rm young}$; see \autoref{sec: fesc_vs_prop}) and the total mass of star clusters in their corresponding $\HII$ region. The two bottom panels show the extinction value, $E(B-V)$, and the galactocentric radius of the $\HII$ regions, obtained from the SIGNALS catalogue. We do not find clear evidence of correlation between any of the aforementioned properties and $\f$.
   }
    \label{fig: cluster_statistics}
\end{figure}

We have found above that $\f$ varies systematically with $\Ha$ luminosity (\autoref{fig: Fesc_Bayesian}), with less luminous $\HII$ regions having systematically higher $\f$. 
Interestingly, this result differs from the one obtained by \citet{2012ApJ...755...40P} for the Magellanic Clouds; they found that $\HII$ regions with higher $\LHa$ are leakier. This is not too surprising, given that NGC 628 differs from the Magellanic Clouds in many ways (e.g., morphology, star formation rate, metallicity); as such, our result may represent changing ionisation conditions in different galactic environments \citep[e.g., see][for the effects of local environment on $\f$]{2017MNRAS.470.4453R}. The introduction of both LEGUS and SIGNALS surveys is therefore timely and will offer a crucial opportunity to compare observations over a wide range of galactic environments.

It is also interesting to search for systematic correlations between $\f$ and other $\HII$ region or star cluster properties. For example, we might envision that older clusters, which have had more time to clear their environments, might have systematically leakier $\HII$ regions. We might also expect that some of the more luminous regions are less leaky due to a higher proportion of the LyC photons being absorbed by dust \citep[e.g., see][]{1986PASP...98..995M, 1972ApJ...177L..69P, 2004ApJ...608..282A}, or that the efficiency of LyC escape might be influenced by metallicity \citep[e.g., see][]{2019MNRAS.486.2215K}, which in turns varies with galactocentric radius \citep[e.g., see][]{1983MNRAS.204...53S, 2000A&A...363..537R, 2010ApJ...721L..48K, 2011MNRAS.415..709S, 2012A&A...540A..56P, 2018A&A...609A.102R}. To investigate all these possibilities, in \autoref{fig: cluster_statistics} we plot $\f$ as a function of cluster age, cluster mass, extinction, and galactocentric radius\footnote{For cluster age and mass, we face two complications in assigning single numbers to these quantities: first, \textsc{slug} returns PDFs for these quantities, not single numbers. Second, many $\HII$ regions are associated with multiple clusters. For this purpose, we first take the age and mass of each cluster to be the 50th percentile value returned by \textsc{slug}. We then simply sum the 50th percentile masses to define a total cluster mass, while for age we adopt a ``young'' age estimate set equal to the 25th percentile of the ages of the clusters associated with a given $\HII$ region. We bias the age estimate young in this way because under the hypothesis that low $\f$ regions are caused by gas clearing, what matters are the clusters that have had the least time to clear their gas.}. The figure reveals no clear systematic correlations between $\f$ and any of these properties; though of course, we cannot rule out the presence of such a correlation, especially because our values of both $\f$ and cluster properties have very large uncertainties at the level of individual clusters. Nonetheless, we find no evidence for $\f$ correlating with any cluster / $\HII$ regions properties other than $\LHa$.

\begin{figure*}
    \centering
    \includegraphics[width = .8\textwidth]{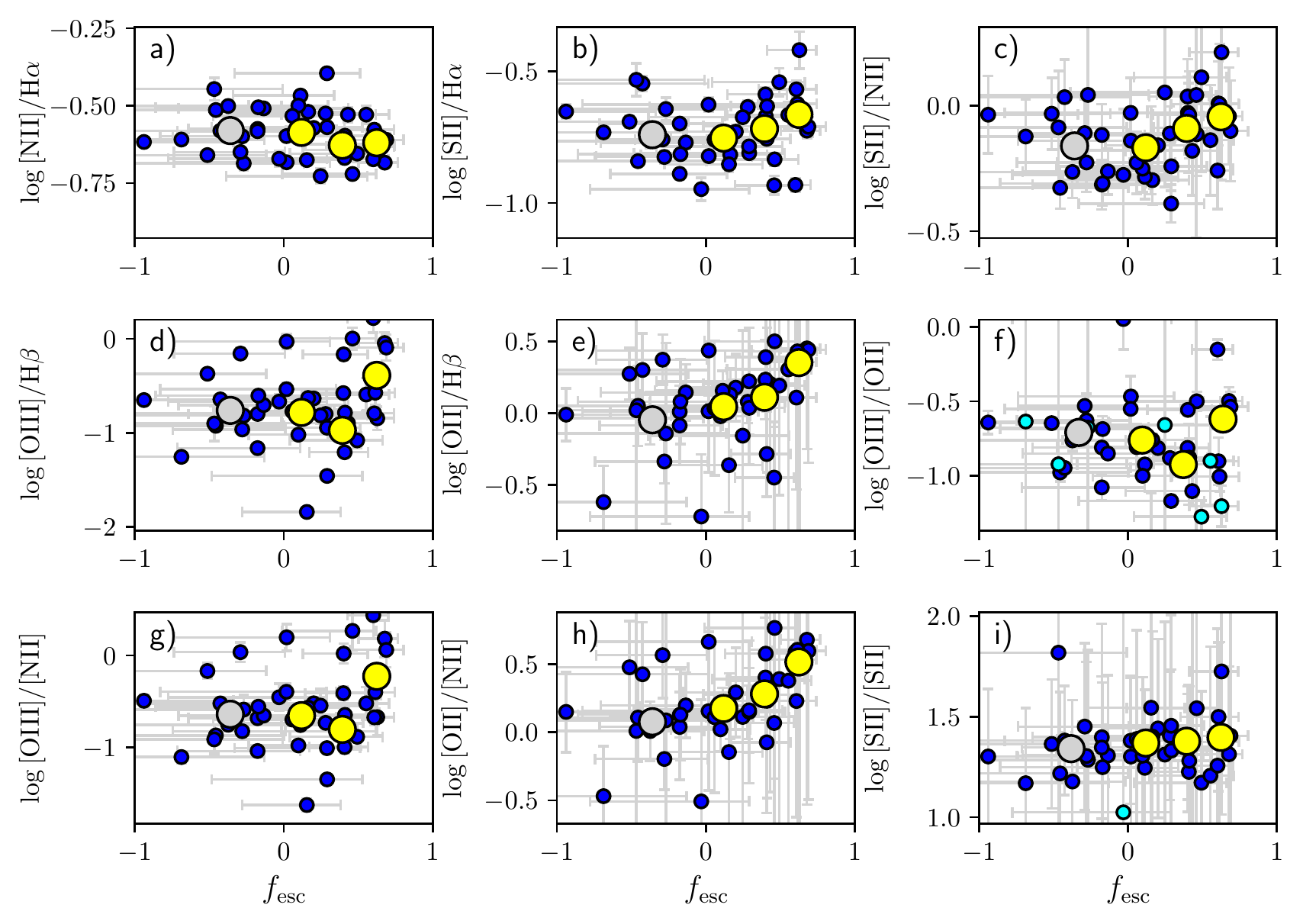}
    \caption{ $\f$ vs extinction-corrected emission line ratios. Blue and cyan points are line ratios with SNR$_{\rm cross}\geq 3$ and SNR$_{\rm cross}<3$ respectively, where SNR$_{\rm cross}$, represented by vertical errorbars, is defined as the line ratio's best SNR \citep[see][for more details]{2018MNRAS.477.4152R}. The horizontal errorbars indicate the 68\% confidence interval of $\f$. Yellow data points represent the mean value obtained by averaging the data in bins of $\f = 0-0.25, 0.25-0.5$ and $0.5-0.75$, using the 50th percentile estimate of $\f$ to assign each $\HII$ region to a bin, and excluding data with SNR$_{\rm cross}<3$. We show the mean and $68\%$ confidence range for each cluster as blue points with horizontal errorbars. In \autoref{tab: lineratio} we present the detailed description of the line ratios as well as the Spearman's rank correlation coefficient, $\rho$, and the corresponding $p-$value.
             }
    \label{fig: lineratio}
\end{figure*}

\begin{table*}
	\centering
	\caption{Emission line ratios vs $\f$ and their corresponding Spearman's rank correlation coefficient $\rho$ and $p-$value (See \autoref{fig: lineratio}).}
	\renewcommand{\arraystretch}{1.8}
	\begin{threeparttable}
	\begin{tabular}{lccclcc}
		\toprule
		\toprule 
        Emission line ratio & $\rho$ & $p-$value\tnote{a} && Emission line ratio & $\rho$ & $p-$value \\
		\midrule
        $a)\;\rm[N\,\textsc{ii}]\lambda6583/\Ha$ & $-0.26$ & 0.0972 && $f)\;\rm[O\,\textsc{iii}]\lambda5007/[O\,\textsc{ii}]\lambda3727$ & -0.02 & 0.9302 \\
        $b)\;\rm[S\,\textsc{ii}]\lambda6716+6731/\Ha$ & 0.14 & 0.3651 && $g)\;\rm[O\,\textsc{iii}]\lambda5007/[N\,\textsc{ii}]\lambda6583$ & 0.18 & 0.2571 \\
        $c)\;\rm[S\,\textsc{ii}]\lambda6716+6731/[N\,\textsc{ii}]\lambda6583$ & 0.27 & 0.0818 && $h)\;\rm[O\,\textsc{ii}]\lambda3727/[N\,\textsc{ii}]\lambda6583$  & 0.52 & 0.0004 \\
        $d)\;\rm[O\,\textsc{iii}]\lambda5007/H\beta$ & 0.17 & 0.2862 && $i)\;\rm[S\,\textsc{ii}]\lambda6716/[S\,\textsc{ii}]\lambda6731$ & 0.14 & 0.3717 \\
        $e)\;\rm[O\,\textsc{ii}]\lambda3727/H\beta$ & 0.48 & 0.0014 &&  & & \\
		\bottomrule
	\end{tabular}
	\begin{tablenotes}
    \item[a]The $p-$value is quoted for a hypothesis test whose null hypothesis is that two sets of data are uncorrelated. The $\rm[O\,\textsc{ii}]\lambda3727/[N\,\textsc{ii}]\lambda6583$ and $\rm[O\,\textsc{ii}]\lambda3727/H\beta$ lines show significant correlation (i.e., $p < 0.005$).
	\end{tablenotes}
	\label{tab: lineratio}
	\end{threeparttable}
\end{table*}

\subsubsection{$\f$ and emission line ratios}
\label{sec: emission}

In \autoref{fig: lineratio}, we plot $\f$ versus these emission line ratios. To quantify possible correlations, we show the Spearman's rank correlation coefficient $\rho$ and the corresponding $p-$value in \autoref{tab: lineratio}. We find that both $\rm[O\,\textsc{ii}]/[N\,\textsc{ii}]$ and $\rm[O\,\textsc{ii}]/H\beta$ correlate with $\f$ at high statistical significance ($p < 0.0055$).\footnote{Note that, since we are testing 9 candidate line ratios for correlation with $\f$, Bonferroni correction implies that in order to conclude at 95\% confidence that \textit{any} of the line ratios are correlated with $\f$, we must obtain a $p$-value $p<0.05/9 = 0.0055$ for any particular line ratio; both $\rm[O\,\textsc{ii}]/[N\,\textsc{ii}]$ and $\rm[O\,\textsc{ii}]/H\beta$ reach this significance level.} The physical origin of the correlation is unclear, though we note the possibility that it could relate to either metal abundance or ionisation parameter \citep[e.g., see][]{2019ApJ...887...80K, 2022ApJ...929..118G, 2023MNRAS.522.2369S}.
Further progress on this question likely requires improved stellar models in order to obtain better constraints on cluster properties \citep[e.g.,][]{2021ApJ...908..241G}, coupled with theoretical studies and forward-modelling to predict observable correlations between LyC escape and other line diagnostics. For instance, \citet[][]{2020MNRAS.496..339P} used a combination of stellar feedback code \textsc{warpfield} \citep[][]{2017MNRAS.470.4453R, 2019MNRAS.483.2547R}, $\HII$ region code \textsc{cloudy} \citep{2017RMxAA..53..385F}, radiative transfer code \textsc{polaris} \citep{2016A&A...593A..87R} and SITELLE observations of $\HII$ regions in NGC~628 \citep{2018MNRAS.477.4152R} to directly compare between emission lines from feedback models and observations.

\section{Conclusion}
\label{sec: conclusion}
 
We present in this work a pilot study of the association between spatially resolved $\HII$ regions and star clusters in the nearby spiral galaxy NGC 628. We combine 334 $\HII$ regions from the SIGNALS survey \citep{2018MNRAS.477.4152R, 2019MNRAS.489.5530R} and 1253 star clusters from the LEGUS survey \citep{2015AJ....149...51C, 2015ApJ...815...93G}, which we link based on their spatial overlap. In the combined catalogue, we find 49\% of $\HII$ regions lack ionising star clusters. 

We begin in \autoref{sec:completeness} by analysing six possible explanations for this phenomenon (see \autoref{tab: orphan_overview} for an overview). We consider and reject explanations that the orphan $\HII$ regions are caused by errors in the astrometry, runaway stars, OB associations, or errors in how SIGNALS segments $\HII$ regions. Ultimately, we find that the dominant contributor to the orphan $\HII$ regions is the incompleteness of the LEGUS catalogue, with the major component of this being clusters that are able to create $\HII$ regions detectable by SIGNALS but are too faint to be classified by LEGUS. A subdominant contributor is that the LEGUS visual classification mistakenly identifies some true clusters as Class~4 (non-clusters).

In the second part of this work (\autoref{sec:escape_fraction}) we attempt to infer the escape fraction of ionising photons from $\HII$ regions by determining the ionising luminosity of star clusters, $\QH$ from a combination of photometric data and stochastic stellar population model \citep[\textsc{slug};][]{2012ApJ...745..145D, 2014MNRAS.444.3275D, 2015MNRAS.452.1447K}, and comparing this to the ionising luminosity required to power the matched $\HII$ regions (see \autoref{fig: LHa_QH0_wCons}). Overall, we find $\f = 0.09^{+0.06}_{-0.06}$ across our population, with $\f = 0.56^{+0.08}_{-0.14}$ and $0.06^{+0.06}_{-0.06}$ for low- and high-$\LHa$ bins respectively (see \autoref{fig: Fesc_Bayesian}). We also find that the escape fraction correlates at high statistical significance ($p < 0.0055$) with metallicity tracers, such as $\rm[O\,\textsc{ii}]/[N\,\textsc{ii}]$ and $\rm[O\,\textsc{ii}]/H\beta$.

Overall, we have demonstrated the use of \textsc{slug} as a powerful tool to study the demographics of star cluster populations at extragalactic distances with consideration of the effect of stochastic sampling of the stellar IMF. Combining with the LEGUS and SIGNALS surveys, this allows us to better understand the interaction between stars and their surrounding interstellar medium with high accuracy where most conventional methods cannot.

\section*{Acknowledgements}
We are grateful for the enlightening discussions and valuable comments on this work by an anonymous referee that improved the scientific outcome and quality of the paper. The authors appreciate the useful discussions and suggestions on this work by J.\ Chisholm, N.\ V.\ Asari, G.\ Stasinska, C.\ Morisset, and I.\ P\'erez.
KG is supported by the Australian Research Council through the Discovery Early Career Researcher Award (DECRA) Fellowship DE220100766 funded by the Australian Government. 
This work is supported by the Australian Research Council Centre of Excellence for All Sky Astrophysics in 3 Dimensions (ASTRO~3D), through project number CE170100013. 
MRK acknowledges support from the Australian Research Council through its Future Fellowship funding scheme, award FT180100375.
MM acknowledges the support of the Swedish Research Council, Vetenskapsr{\aa}det (grant 2019-00502). 
RSK is thankful for financial support from the European Research Council in the ERC Synergy Grant `ECOGAL' (project ID 855130), from the German Science Foundation (DFG) via the Collaborative Research Center ’The Milky Way System’ (SFB 881, Funding-ID 138713538, subprojects A1, B1, B2, B8) and from the Heidelberg Cluster of Excellence `STRUCTURES' (EXC 2181 - 390900948). RSK and JWT also acknowledge funding from the German Space Agency (DLR) and the Federal Ministry for Economic Affairs and Climate Action (BMWK) in project `MAINN' (grant number 50OO2206). The group in Heidelberg also thanks for computing resources provided by the Ministry of Science, Research and the Arts (MWK) of the State of Baden-W\"{u}rttemberg through bwHPC and DFG through grant INST 35/1134-1 FUGG and for data storage at SDS@hd through grant INST 35/1314-1 FUGG.  
MF is thankful for financial support from the European Research Council (ERC) under the European Union's Horizon 2020 research and innovation programme (grant agreement No 757535) and by Fondazione Cariplo (grant No 2018-2329).
LC acknowledges financial support from ANID/Fondecyt Regular project 1210992.
JW acknowledges support by the NSFC grants U1831205 and 12033004.
This research is based on observations made with the NASA/ESA Hubble Space Telescope, obtained at the Space Telescope Science Institute, which is operated by the Association of Universities for Research in Astronomy, under NASA Contract NAS 5–26555. These observations are associated with Program 13364. Support for Program 13364 was provided by NASA through a grant from the Space Telescope Science Institute.
The SIGNALS observations were obtained with SITELLE, a joint project between Universit\'{e} Laval, ABB-Bomem, Universit \'{e} de Montreal, and the CFHT, with funding support from the Canada Foundation for Innovation (CFI), the National Sciences and Engineering Research Council of Canada (NSERC), Fonds de Recherche du Quebec – Nature et Technologies (FRQNT), and CFHT.
The authors wish to recognize and acknowledge the very significant cultural role that the summit of Maunakea has always had within the indigenous Hawaiian community. The authors are most grateful to have the opportunity to conduct observations from this mountain with the CFHT.

\section*{Data Availability}
The data and the source code underlying this article will be available at \url{https://github.com/JiaWeiTeh/LEGUS-SIGNALS-NGC628}.



\bibliographystyle{mnras}
\bibliography{reference}


\section*{Affiliations}
\noindent
{\it
$^{1}$Universit\"{a}t Heidelberg, Zentrum f\"{u}r Astronomie, Institut f\"{u}r Theoretische Astrophysik, Albert-Ueberle-Stra{\ss}e 2, D-69120 Heidelberg, Germany\\
$^{2}$Research School of Astronomy and Astrophysics, Australian National University, Canberra, ACT 2601, Australia\\
$^{3}$ARC Centre of Excellence for All Sky Astrophysics in 3 Dimensions (ASTRO 3D), Australia\\
$^{4}$Department of Astronomy, University of Massachusetts, Amherst, MA 01003, USA\\
$^{5}$Canada–France–Hawaii Telescope, Kamuela, HI, 96743, USA\\
$^{6}$Department of Physics and Astronomy, University of Hawaii at Hilo, Hilo, HI, 96720-4091, USA\\
$^{7}$Département de Physique, Université de Montréal, Succ. Centre-Ville, Montréal, Québec, H3C 3J7, Canada\\
$^{8}$Centre de Recherche en Astrophysique du Québec (CRAQ), Québec, QC, G1V 0A6, Canada\\
$^{9}$The Oskar Klein Centre, Department of Astronomy, Stockholm University, AlbaNova, SE-10691 Stockholm, Sweden\\
$^{10}$Steward Observatory, University of Arizona, Tucson, AZ  85721-0065, USA\\
$^{11}$George P. and Cynthia W. Mitchell Institute for Fundamental Physics \& Astronomy, Texas A\&M University, College Station, TX 77843-4242, USA \\
$^{12}$Astronomisches Rechen-Institut, Zentrum f\"ur Astronomie der Universit\"at Heidelberg, M\"onchhofstr.\ 12--14, 69120 Heidelberg, Germany\\
$^{13}$Caltech/IPAC, 1200 E. California Boulevard, Pasadena, CA 91125, USA\\
$^{14}$Observatoire de Paris, LERMA, Coll\`ege de France,  PSL Univ., CNRS, Sorbonne Univ., F-75014, Paris, France\\
$^{15}$Observatoire de Gen\`eve, Universit\'e de Gen\`eve, Versoix, Switzerland\\
$^{16}$Universit\"{a}t Heidelberg, Interdisziplin\"{a}res Zentrum f\"{u}r Wissenschaftliches Rechnen, Im Neuenheimer Feld 205, D-69120 Heidelberg, Germany\\
$^{17}$Instituto de Astrofisica de Andalucia (CSIC), Glorieta de la Astronomia 
s/n, 18008 Granada, Spain\\
$^{18}$Dipartimento di Fisica G. Occhialini, Universit\`a degli Studi di Milano-Bicocca, Piazza della Scienza 3, 20126 Milano, Italy\\
$^{19}$INAF - Osservatorio Astronomico di Trieste, via G. B. Tiepolo 11, 34143 Trieste, Italy\\
$^{20}$Centre for Extragalactic Astronomy, Department of Physics, Durham University, South Road,  Durham DH1 3LE, United Kingdom\\
$^{21}$Institute for Computational Cosmology, Department of Physics, University of Durham, South Road, Durham DH1 3LE, United Kingdom\\
$^{22}$Space Telescope Science Institute, 3700 San Martin Drive, Baltimore, MD 21218, USA\\
$^{23}$Instituto de Astrofisica, Universidad Andres Bello, Fernandez Concha 700, Las Condes, Santiago RM, Chile\\
$^{24}$Department of Astronomy, Xiamen University, Xiamen, Fujian 361005, China\\
$^{25}$Leibniz-Institut für Astrophysik Potsdam (AIP), An der Sternwarte 16, 14482 Potsdam, Germany\\
$^{26}$Aix Marseille Univ, CNRS, CNES, LAM, Marseille, France\\
$^{27}$Visiting Fellow, Harvard-Smithsonian Center for Astrophysics, 60 Garden Street, Cambridge, MA 02138, USA   \\
}


\appendix

\section{The effect of using alternative conditions on $\f$}
\label{ap: alternative_libraries}

In the main text, we make three methodological choices that warrant further exploration. These are the choice of Geneva tracks for stellar evolution, the inclusion of Class~4 objects as potential ionising sources, and not imposing a prior that $\f$ be non-negative. In this appendix we explore the consequences of making different choices. To this end, we repeat the analysis presented in \autoref{sec: individual} and \autoref{sec: population} with the following combinations of alternative choices: (1) Geneva tracks with no prior on $\f$, but only using Class~1-3 sources, (2) Padova tracks with no prior on $\f$ and using Class~1-4 sources, (3) Padova tracks with no prior on $\f$ and using only Class~1-3 sources, and (4) Geneva tracks using Class~1-4 sources, but adopting a prior that $\f\geq 0$.

We report the marginal posterior PDFs for $\f$ produced by each of these alternative approaches in \autoref{tab: alternative_models}. If we focus on the population-level analysis, which is more reliable and has smaller uncertainties, we find that the only methodological choice that produces a difference in the outcome that is larger than the uncertainties is the use of Padova rather than Geneva tracks. We prefer the Geneva tracks in our analysis because doing so produces more data points with smaller uncertainty compared to Padova tracks, which is not surprising as the former are more carefully tuned to model observations of young and massive stars. All other methodological choices produce shifts that are smaller than the uncertainties. For example, if we exclude Class 4 sources, escape fractions shift to somewhat lower values, reaching median $\f<0$ in the high luminosity group, which is certainly suggestive that Class~4 sources contain true clusters that contribute a significant amount of ionising radiation. However, the uncertainty intervals we derive with and without Class 4 nonetheless slightly overlap.

\begin{figure}
    \centering
    \includegraphics[width = \columnwidth]{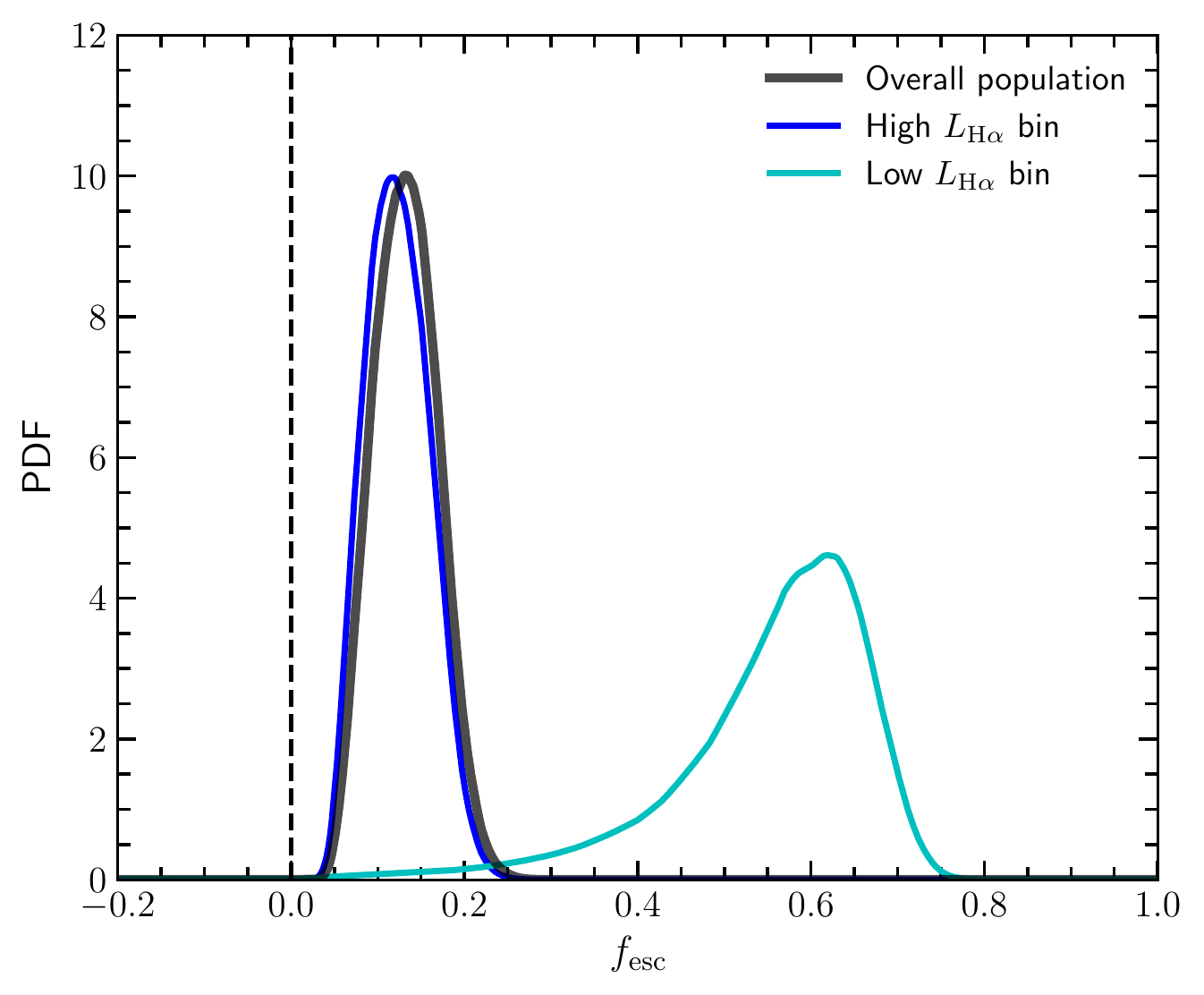}
    \caption{Same as \autoref{fig: Fesc_Bayesian}, except here we impose that $\f\geq 0$ as a complementary result to that of \autoref{sec: f_prob_dist}. We observe similar qualitative behaviour in both results.}
    \label{fig: Fesc_Bayesian_positive}
\end{figure}

\begin{figure*}
    \centering
    \includegraphics[width = .8\textwidth]{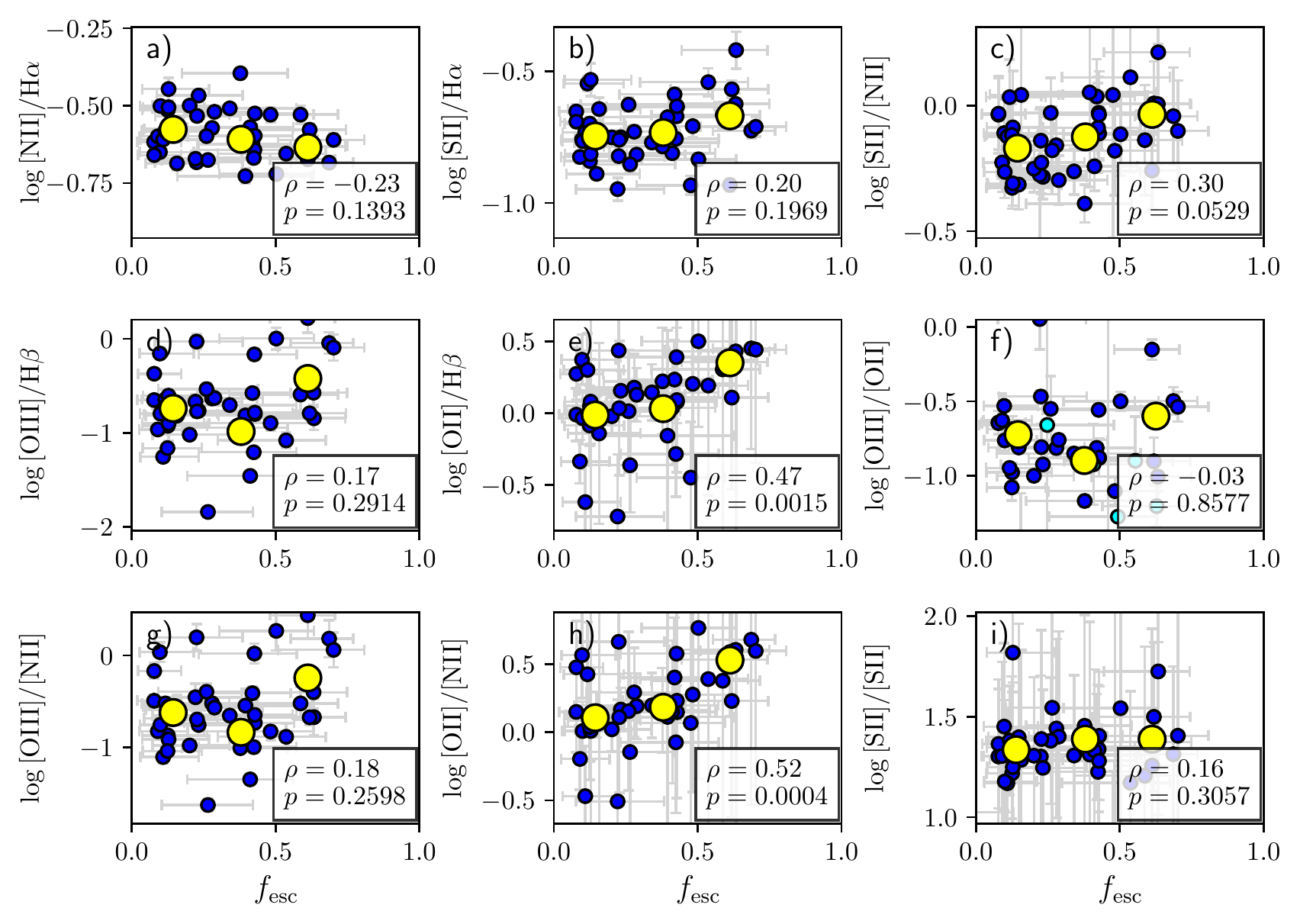}
    \caption{Same as \autoref{fig: lineratio}, except here we impose that $\f\geq 0$ as a complementary result to that of \autoref{sec: f_prob_dist}. We also show the Spearman's rank correlation coefficient $\rho$ and the corresponding $p$--value. We note the correlation in the line ratios $\rm[O\,\textsc{ii}]/[N\,\textsc{ii}]$ and $\rm[O\,\textsc{ii}]/H\beta$, as similarly observed in \autoref{fig: lineratio}.}
    \label{fig: lineratio_positive}
\end{figure*}

It is perhaps more surprising that imposing a prior $\f\geq 0$ does not have larger effects. While there certainly is a modest increase in median $\f$ for the high-luminosity part of the sample, this is again comparable in size to the uncertainties. To explore further why this should be, we plot the population-level marginal posterior PDF of $\f$ derived with a prior $\f \geq 0$ in \autoref{fig: Fesc_Bayesian_positive}, and show the corresponding relationship of $\f$ with emission line ratios in \autoref{fig: lineratio_positive}. Comparing \autoref{fig: Fesc_Bayesian_positive} with \autoref{fig: Fesc_Bayesian}, it is clear that the effect of this prior is cut off the low $\f$ tail of the high $L_\mathrm{H\alpha}$ bin, but that this does not affect the peak of the distribution, or substantially alter the mean. Similarly, comparing \autoref{fig: lineratio_positive} with \autoref{fig: lineratio}, we see that imposing a prior that $\f\geq 0$ removes the high-uncertainty, low $\f$ part of the posterior distribution, but that this part of the distribution was not responsible for driving the correlations we find between $\f$ and [O~\textsc{ii}]/H$\beta$ or [O~\textsc{ii}]/[N~\textsc{ii}]. Thus the significance level of these correlations is not substantially different than we found without imposing a prior. We therefore conclude that our primary qualitative results are insensitive to whether or not we choose to enforce a positive-only prior of $\f$.

\begin{table*}
	\centering
	\caption{Alternative model parameters and their effect on our analysis. Column (1) indicates which stellar evolution tracks we use for the \textsc{slug} models, column (2) indicates whether we include class 4 LEGUS sources as potential drivers of ionisation, and column (3) indicates whether we impose a prior on $\f$ in our Bayesian analysis.}
	\renewcommand{\arraystretch}{2}
	\begin{threeparttable}
	\begin{tabular}{cccccccccccccc}
		\toprule
		\toprule 
		\multirow[1]{2}{*}{Tracks} & \multirow{2}{*}{Class 4?} &
		\multirow{2}{*}{$\f$ prior?} &
		\multirow{2}{*}{$n$\tnote{a}} &&
		\multicolumn{3}{c}{$\f$ (Individual, \autoref{sec: individual})} && \multicolumn{3}{c}{$\f$ (Population, \autoref{sec: population})} \\\cmidrule{6-8}\cmidrule{10-12}
		& & & & & Low\tnote{b}  & High & Overall & & Low & High & Overall \\
		\midrule
		Geneva\tnote{c} & Y & N & 42 &&$0.34^{+0.31}_{-0.70}$& $-0.07^{+0.47}_{-0.75}$& $0.13^{+0.43}_{-0.76}$  && $0.56^{+0.08}_{-0.14}$&$0.06^{+0.06}_{-0.06}$&$0.09^{+0.06}_{-0.06}$ \\
		Geneva & N & N & 33 && $0.22^{+0.37}_{-0.96}$ & $-0.20^{+0.56}_{-1.13}$ & $0.04^{+0.46}_{-1.02}$ & &  $0.55^{+0.10}_{-0.19}$&$-0.07^{+0.07}_{-0.08}$&$-0.04^{+0.07}_{-0.08}$ \\
		Padova & Y & N & 21 &&$0.12^{+0.42}_{-0.97}$ &$-0.03^{+0.51}_{-1.01}$ & $0.04^{+0.47}_{-1.01}$ & &$0.20^{+0.30}_{-0.51}$&$0.00^{+0.08}_{-0.12}$&$0.02^{+0.08}_{-0.11}$\\
		Padova & N & N &17 &&$0.23^{+0.39}_{-1.02}$ &$-0.37^{+0.65}_{-1.34}$ & $-0.06^{+0.55}_{-1.33}$& &$0.57^{+0.16}_{-0.50}$&$-0.16^{+0.11}_{-0.13}$&$-0.14^{+0.11}_{-0.12}$\\
         Geneva & Y & Y & 42 &&$0.48^{+0.21}_{-0.27}$& $0.30^{+0.24}_{-0.21}$& $0.41^{+0.23}_{-0.27}$  && $0.58^{+0.08}_{-0.13}$&$0.12^{+0.04}_{-0.04}$&$0.13^{+0.04}_{-0.04}$ \\
		Geneva\tnote{d} & Y & N & 37&&$0.39^{+0.30}_{-0.59}$&$0.01^{+0.51}_{-0.67}$&$0.21^{+0.42}_{-0.70}$ && $0.55^{+0.11}_{-0.19}$&$0.15^{+0.05}_{-0.05}$&$0.16^{+0.05}_{-0.05}$ \\
		\bottomrule
	\end{tabular}
	\begin{tablenotes}
	\item[a] Number of samples after removing sources with large uncertainties and sources in coloured regions (see \autoref{sec: individual}).
	\\\item[b] Values quoted for low- and high-$\LHa$ bins, and the overall population (see \autoref{sec: f_prob_dist}).
	\\\item[c] Parameters used in the main text. 
	\\\item[d] Same parameter choices as in the main text, but with a slight modification of our approach to associating clusters and $\HII$ regions, as described in \aref{ap: peculiar}.
	\end{tablenotes}
	\label{tab: alternative_models}
	\end{threeparttable}
\end{table*}

\section{Visualising the distribution of $\QH$}
\label{ap: visualise_QH0}
\begin{figure*}
    \centering
    \includegraphics[width = .8\textwidth]{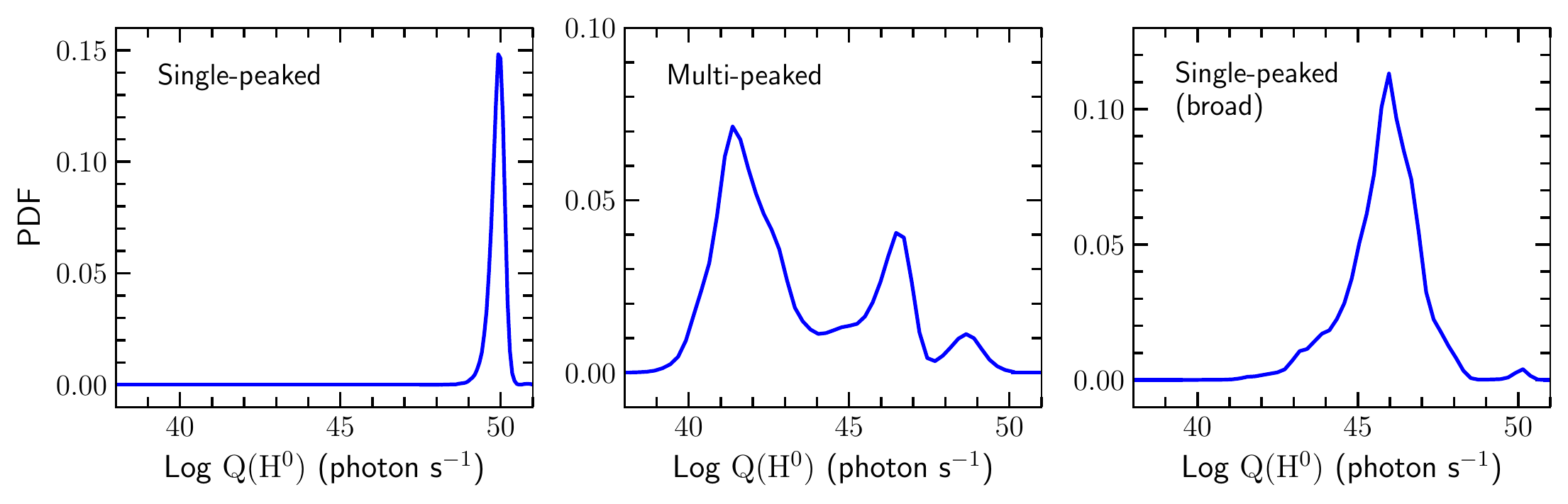}
    \caption{Examples of the posterior PDF of $\QH$. From right to left panel: (1) sharp, single-peaked profile; (2) multi-peaked profile; and (3) broad ($\sigma > 0.5$ dex), single-peaked profile (see \autoref{sec: cluster_dist}).}
    \label{fig: visualise_QH0}
\end{figure*}

In \autoref{sec: cluster_dist} we discussed the distributions of cluster ionising luminosity, $\QH$, under the effect of stochastic IMF sampling, and how they affect our analysis. To illustrate further, in \autoref{fig: visualise_QH0} we show three common distributions of $\QH$ found in this work: (1) sharp, single-peaked profile; (2) multi-peaked profile; and (3) broad ($\sigma > 0.5$ dex), single-peaked profile.

\section{Possible explanations for peculiar data points and their implications}
\label{ap: peculiar}
\begin{figure}
    \centering
    \includegraphics[width = \columnwidth]{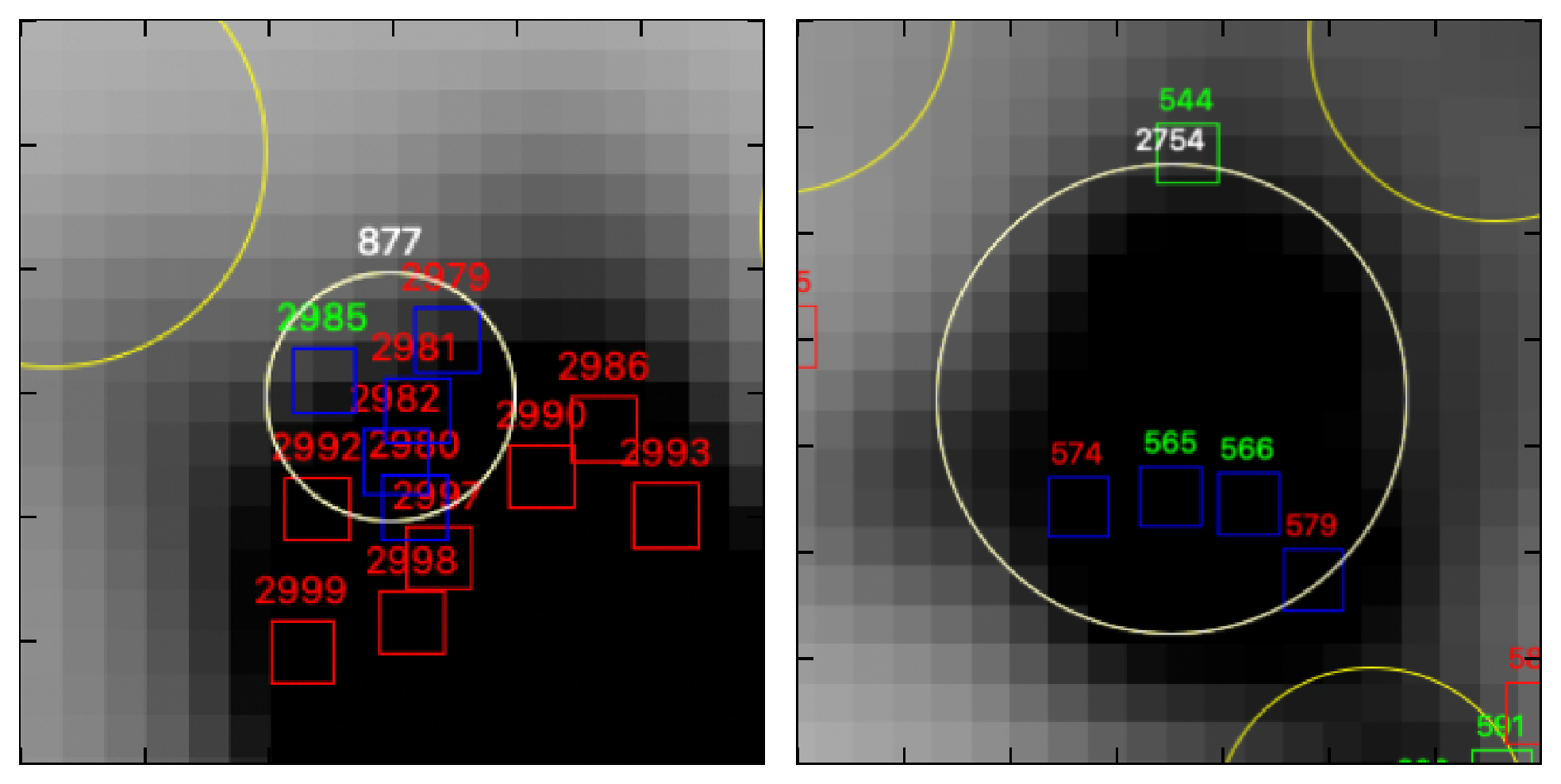}
    \caption{Visual inspection of the two data points in blue regions (see \autoref{fig: LHa_QH0_wCons}), overlaid on $\Ha$ map obtained from the SIGNALS survey. Each figure outlines the target $\HII$ region (in white circle) and its corresponding star cluster associations (in blue squares). LEGUS sources that are not associated with the target regions are either outlined with red squares (class 1, 2, 3), or with green squares (class 4), whereas the surrounding SIGNALS $\HII$ regions are outlined in yellow circles. Additionally, the IDs are shown above every source in white (target $\HII$ region), red (class 1, 2, 3 LEGUS sources) and yellow (class 4 LEGUS sources).}
    \label{fig: top2}
\end{figure}

\begin{figure}
    \centering
    \includegraphics[width = \columnwidth]{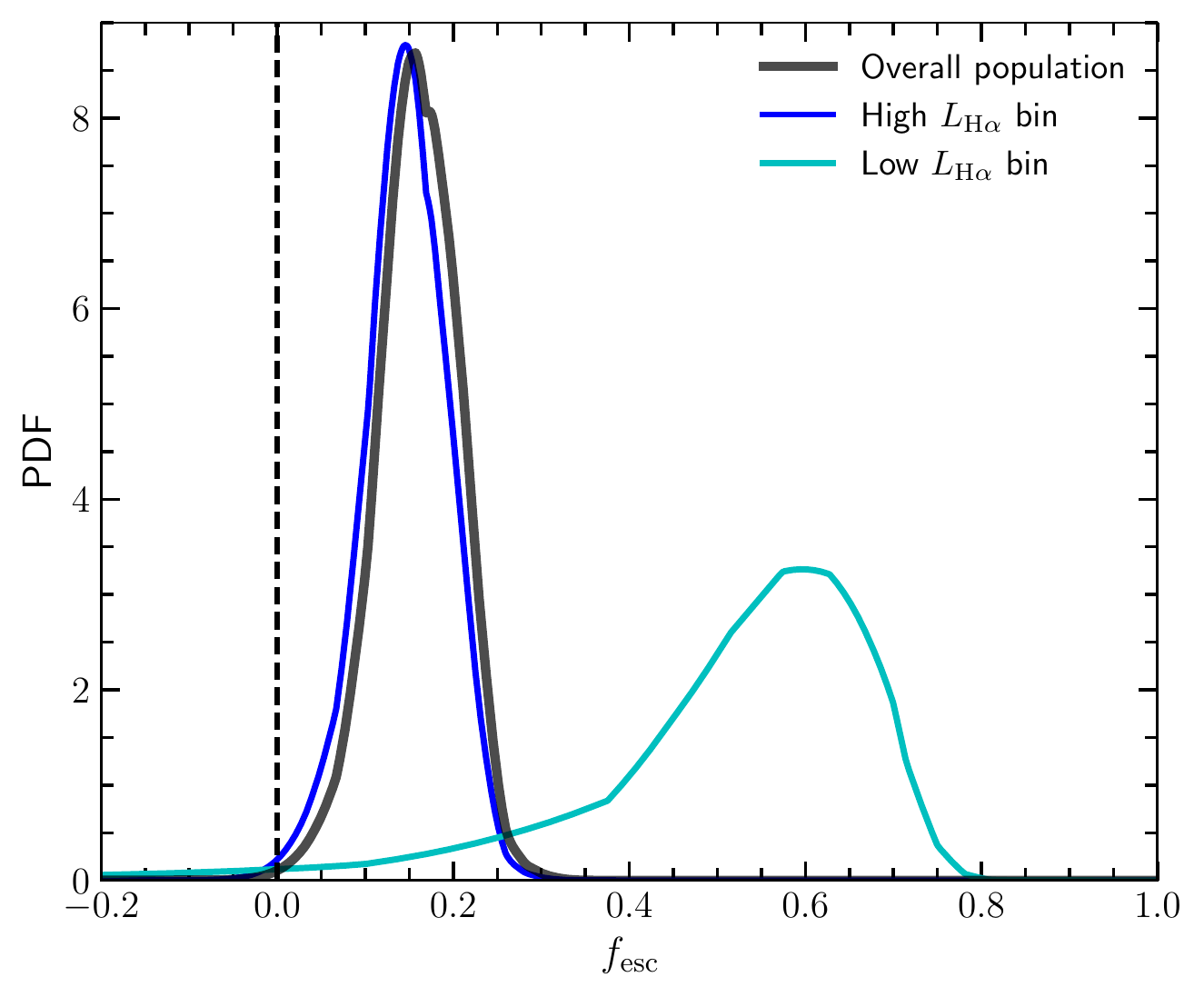}
    \caption{Same as \autoref{fig: Fesc_Bayesian}, except here we include clusters within $1.5R$ of $\HII$ regions, following our reasoning in \aref{ap: peculiar}.}
    \label{fig: a15_Fesc_Bayesian}
\end{figure}

\autoref{fig: LHa_QH0_wCons} shows two sets of data points (in green and blue regions) that are well within the forbidden zone and are excluded from our analysis. We argue that data points in the green region are either a result of false associations, or that we are missing ionising clusters due to the completeness of the LEGUS survey. These two arguments, however, cannot fully explain the two data points in the blue region for the following reasons. First, it is highly unlikely that the detected $\HII$ regions and star clusters -- both at the high luminosity end -- are chance alignments. Second, as illustrated by \autoref{fig: SLUG_completeness}, the most luminous $\HII$ regions are consistently powered by star clusters that are classified by the LEGUS survey. This suggests that the peculiar points could be a consequence of missed diffuse ionising sources, as we have discussed in \autoref{sec: diffuse}.

To further investigate these two data points, we visually examine their location by overlaying LEGUS and SIGNALS data onto the SIGNALS $\Ha$ map (see \autoref{fig: top2}). An interesting feature presented in both sub-figures is that there are LEGUS sources that clearly located within an $\Ha$ detection but are not placed in an $\HII$ region. Regardless, this analysis shows that it is plausible that the $\HII$ region radius reported by SIGNALS can cause an underestimation of the area of influence, thus missing potential ionising sources. To study this effect, we repeat the analysis in \autoref{sec:escape_fraction}, except this time we modify our approach as follows: for each $\HII$ region in the truncated sample, we consider additional ionising clusters if the projected distance, $D$, between the central positions is $\leq 1.5 R$, where $R$ is the radius of $\HII$ region specified by the SIGNALS catalogue. We show our results of $\f$ using this approach in \autoref{tab: alternative_models} and \autoref{fig: a15_Fesc_Bayesian}, and report that this new approach leads to similar qualitative conclusions.

\section{Star cluster and H \textsc{ii} region catalogues}
\label{ap: catalogues}
We present in \autoref{tab: starcluster_catalogue} and \autoref{tab: h2region_catalogue} catalogue entries of star clusters and $\HII$ regions used for this paper.

\begin{table*}
	\centering
	\caption{The LEGUS star cluster catalogue in the overlapping FoV for NGC~628 \citep[see][]{2015ApJ...815...93G,2017ApJ...841..131A}, showing selected columns and rows.
 (1) ID$_{\rm LEGUS}$: Identification number,
 (2) RA: right ascension,
 (3) DEC: declination,
 (4) Class: LEGUS assigned class,
 (6) ID$_{\rm SIGNALS}$: Identification number of associated $\HII$ region(s) in the SIGNALS catalogue.
}
	\renewcommand{\arraystretch}{1.7}
	\begin{threeparttable}
	\begin{tabular}{cccccc}
		\toprule
		\toprule 
		ID$_{\rm LEGUS}$ & RA (deg) & DEC (deg) & Class & ID$_{\rm SIGNALS}$ \\
		\midrule
        73 & 24.1742 & 15.8082 & 1 & 3253\\
        86 & 24.1751 & 15.8073 & 3 & 3206\\
        ... & ... & ... & ... & ... \\
        1138 & 24.1748 & 15.7859 & 4 & 2330\\
        1140 & 24.1705 & 15.7858 & 4 & 2323, 2359\\
        1141 & 24.1553 & 15.7858 & 3 & -\\
        ... & ... & ... & ... & ... \\
        3671 & 24.2071 & 15.7407 & 2 & -\\
		\bottomrule
	\end{tabular}
	\label{tab: starcluster_catalogue}
	\end{threeparttable}
\end{table*}

\begin{table*}
	\centering
	\caption{The SIGNALS $\HII$ region catalogue in the overlapping FoV for NGC~628 \citep[see][]{2018MNRAS.477.4152R}, showing selected columns and rows.
 (1) ID$_{\rm SIGNALS}$: Identification number,
 (2) RA: right ascension,
 (3) DEC: declination,
 (4) $\LHa$: $\Ha$ luminosity of $\HII$ region,
 (5) $r$: radius of $\HII$ region,
 (6) ID$_{\rm LEGUS}$: Identification number of associated star cluster(s) in the LEGUS catalogue.
}
	\renewcommand{\arraystretch}{1.7}
	\begin{threeparttable}
	\begin{tabular}{cccccc}
		\toprule
		\toprule 
		ID$_{\rm SIGNALS}$ & RA (deg) & DEC (deg) & $\LHa$ (erg s$^{-1}$) & $r$ (pc) & ID$_{\rm LEGUS}$ \\
		\midrule
		397 & 24.2082 & 15.7410 & 1.233 $\times 10^{36}$ & 68.0 & -\\
        414 & 24.2068 & 15.7414 & 1.882 $\times 10^{37}$ & 81.0 & -\\
        ... & ... & ... & ... & ... & ... \\
        2013 & 24.1458 & 15.7805 & 1.656 $\times 10^{37}$ & 107.0 & 1669, 1680\\
        2016 & 24.1777 & 15.7812 & 6.339 $\times 10^{37}$ & 44.3 & 1570, 1571, 1605\\
        2017 & 24.1681 & 15.7810 & 3.446 $\times 10^{37}$ & 44.3 & 1619, 1628 \\
        ... & ... & ... & ... & ... & ... \\
        3253 & 24.1740 & 15.8082 & 5.577 $\times 10^{36}$ & 89.0 & 73\\
    \bottomrule
	\end{tabular}
	\label{tab: h2region_catalogue}
	\end{threeparttable}
\end{table*}


\bsp	
\label{lastpage}
\end{document}